\documentclass[10pt]{article}
\usepackage[left=1.5cm,right=1.5cm]{geometry}
\usepackage[numbers,square]{natbib}
\usepackage[T1]{fontenc}
\usepackage{tikz-cd}
\usepackage[british]{babel}
\usepackage{layout}
\usepackage{tcolorbox} 
\usepackage{amsmath}
\usepackage{amssymb,amstext, amsthm, simplewick, amsfonts}
\usepackage{framed}
\usepackage{pgfplots}
\usepackage{amsfonts}
\usepackage{color} 
\usepackage{braket}
\usepackage{multicol} 
\usepackage{epsfig}
\usepackage{cancel}
\usepackage{caption}
\usepackage{epsf}
\usepackage{feynmf} 
\usepackage{frontespizio}
\usepackage{rotating}
\usepackage{subfigure}
\usepackage{pstricks}
\usepackage{type1ec}
\usepackage{lettrine}
\usepackage{bbold}
\usepackage{calligra}
\usepackage{tikz}
\usepackage{float}
\usepackage{subfigure}
\usepackage{slashed}
\usepackage{mathrsfs}
\usepackage{curve2e}
\usepackage{setspace}
\usepackage{indentfirst}
\usepackage{emptypage}
\usepackage{relsize}
\usepackage{mathrsfs}
\usepackage{stackengine}
\usepackage{calc}
\usepackage{hyperref}
\hypersetup{
	colorlinks,
	citecolor=green,
	filecolor=black,
	linkcolor=blue,
	urlcolor=black
}
\newcommand{\3}[1]{C_{
		\ifthenelse{\equal{\ThreePt}{\empty}}{#1}{
			\ifthenelse{\equal{#1}{\empty}}{\ThreePt}{\ThreePt,#1}}}}
\newcommand{\redef}[1]{{C'}_{
		\ifthenelse{\equal{\ThreePt}{\empty}}{#1}{
			\ifthenelse{\equal{#1}{\empty}}{\ThreePt}{\ThreePt,#1}}}}
\newcommand{\ren}[1]{C_{
		\ifthenelse{\equal{\ThreePt}{\empty}}{#1}{
			\ifthenelse{\equal{#1}{\empty}}{\ThreePt}{\ThreePt,#1}}}}
\newcommand{\sd}[1]{D_{
		\ifthenelse{\equal{\ThreePt}{\empty}}{#1}{
			\ifthenelse{\equal{#1}{\empty}}{\ThreePt}{\ThreePt,#1}}}}

\numberwithin{equation}{section} 
\makeatletter
\@addtoreset{equation}{section}
\newcommand{\bea}{\begin{eqnarray}}
\newcommand{\eea}{\end{eqnarray}}
\makeatother
\newcommand{\beqa}{\begin{eqnarray}}
	\newcommand{\eeqa}{\end{eqnarray}}

\newcommand{\nn}{\nonumber}

      \let\l=\lambda  \let\m=\mu
\let\n=\nu           \let\p=\pi

\newcommand{\bann}{\begin{eqnarray*}}
\newcommand{\eann}{\end{eqnarray*}}

\newcommand{\bmi}{\begin{minipage}}
\newcommand{\emi}{\end{minipage}}

  \let\D=\Delta   
           
    \let\Y=\Psi

\newcommand{\be}{\begin{equation}}
	\newcommand{\ee}{\end{equation}}

\renewcommand{\Im}{\textrm{Im}}
\newcommand{\beq}{\begin{equation}}
	\newcommand{\eeq}{\end{equation}}
\newcommand{\figref}[1]{Fig.~\ref{#1}}			
\newcommand{\secref}[1]{Section~\ref{#1}}		

\newcommand{\ThreePt}{\empty}
\usepackage{accents}
\makeatletter
\newcommand{\xLine}[2][]{\ext@arrow 0359\Rightarrowfill@{#1}{#2}}
\makeatother
\xdefinecolor{darkgreen}{RGB}{102, 204, 70}
\xdefinecolor{darkblue}{RGB}{0, 0, 153}
\usepackage{amssymb}
\usepackage{pifont}

\begin{document}

\begin{center} 
{\bf \Large Topological Sum Rules and Spectral Flows \\}
\vspace{0.3cm}
{\bf \Large of Chiral and Gravitational Axion-like Interactions }
\vspace{0.3cm}

\vspace{1.5cm}
{\bf $^{(1,2)}$Claudio Corian\`{o}, $^{(1)}$Stefano Lionetti, $^{(1)}$Dario Melle }
\vspace{1cm}

{\it $^{(1)}$Dipartimento di Matematica e Fisica, 
Universit\`{a} del Salento and \\ INFN-Lecce, Via Arnesano, 73100 Lecce, Italy\\
National Center for HPC, Big Data and Quantum Computing\\}
{\it $^{(2)}$CNR-Nanotec, Lecce \\} 

\vspace{1cm}

\vspace{.5cm}
\begin{abstract}
We examine the structure of off-shell effective actions arising from chiral and gravitational anomalies, focusing on the $\langle JJJ_A\rangle $ (vector/vector/axial-vector) and $\langle TTJ_A \rangle $ (stress-energy tensors/axial-vector) correlators, relevant in the analysis of  anomaly-driven interactions in perturbation theory. 
Our approach employs both perturbative methods and techniques in conformal field theory within momentum space, extended to encompass chiral anomalies.
The analysis centers on the presence of both particle poles and anomaly poles within these interactions, characterizing their behavior both in the conformal and non conformal limits. 
Through explicit computations, we show that universal sum rules in the longitudinal sector regulate these interactions for all kinematic conditions, extending previous analysis. We discuss the resulting spectral flow, an “area law” of the absorptive part of the anomaly form factors as one moves away from or returns to the conformal point. These features are absent in the local effective action of anomaly interactions, commonly used in the description of axion-like particles. The spectral densities in both cases are shown to be self-similar.
Our results further show that anomaly poles correspond to true particle poles only in the conformal limit, when the interaction describes a massless S-matrix process supported on a null-surface. Under these conditions, they can be effectively described by two kinetically mixed pseudoscalar fields propagating on the light-cone. These studies find application in polarized deeply inelastic scattering, axion-like dark matter and analogue systems such as topological materials. 
\end{abstract}
\end{center}

\newpage

\newpage

\section{Introduction}
A direct approach to studying chiral and gravitational anomalies is through perturbative analysis of the 1PI effective action, particularly the axial-vector/vector/vector ($AVV$ or $\langle JJJ_A \rangle$) and axial-vector/stress-energy/stress-energy ($ATT$ or $\langle TTJ_A \rangle$) correlators. Here, \( J_A \) denotes the axial-vector current, \( J \) the vector currents, and \( T \) the stress-energy tensor, which couples to gravity. The $ATT$ vertex extends anomaly considerations into the gravitational sector.\\
Axial-vector currents may arise from chiral fermions in external fields or through interactions with gravitational waves \cite{delRio:2020cmv}. If the early Universe experienced a conformal phase, such anomaly-induced interactions could enhance fermionic chiral asymmetries. In the $ATT$ case, gravitational waves may further amplify chiral chemical potentials \cite{delRio:2021bnl}, impacting primordial magneto-hydrodynamics (MHD) and the evolution of cosmic magnetic fields \cite{Boyarsky:2011uy, Boyarsky:2015faa, Brandenburg:2017rcb}.\\
In a conformal phase, the three-point functions involving conserved currents and the stress-energy tensor are uniquely determined—up to an overall constant—by conformal symmetry. When chiral anomalies are present, these correlators are likewise uniquely fixed, again modulo an overall constant, by simultaneously imposing conformal constraints and diffeomorphism invariance. Remarkably, the full structure of these solutions can be reproduced entirely within ordinary free field theory realizations.\\
The study of the effective action associated with these interactions—manifesting in the ultraviolet (UV) regime as the exchange of intermediate states in the form of anomaly poles in the conformal limit—is characterized by the emergence of sum rules rather than of conventional asymptotic states, saturated by spectral densities which are simply proportional to $\delta$ functions. These intermediate configurations can be interpreted as ordinary particles only under specific kinematic conditions, namely when the dynamics are constrained to the light cone. In other words, these intermediate states are not ordinary particle states.\\
A related phenomenon occurs in Quantum Chromodynamics (QCD) at small values of the Bjorken variable 
$x$, where the so-called perturbative pomeron emerges. Unlike ordinary particles, the perturbative pomeron cannot be associated with a well-defined asymptotic state in the conventional sense. \\
The purpose of this work is to investigate these features for the two correlators discussed above, exploring their full kinematical structure beyond the conformal limit. This is achieved by including perturbative effects due to fermion masses and demonstrating the existence of mass-independent sum rules in all cases. In a related work, we are going to show that a similar result holds for conformal correlators such as the $TJJ$, involving the stress energy tensor and non-abelian vector currents. \\
 A key result of our analysis is that the local effective action for an axion-like particle—originating in the UV through chiral anomaly interactions in a conformal setting—necessarily requires, for its extension at low energy, the inclusion of a nonperturbative intermediate scale in order to be consistently defined as a local action for an asymptotic axion. The only requirement  in the UV is described by a conformal interaction. This finding applies not only to standard chiral anomaly interactions but also to gravitational anomalies. Our detailed investigation around the conformal point demonstrates that the presence of a sum rule is essential to reach such a conclusion.  \\
Our analysis focuses on both correlators. In the conformal limit, the framework is based on momentum-space conformal field theory (CFT$_p$), where anomaly poles in the longitudinal sector of $J_A$ play a central role in solving the conformal Ward identities for both correlators \cite{Coriano:2013jba, Bzowski:2013sza, Bzowski:2015pba, Coriano:2020ees}. These poles are fundamental to understanding both parity-odd chiral anomalies and parity-even trace anomalies in the conformal limit, and they are intricately linked to the full vertex structure, including the transverse components \cite{Coriano:2023cvf, Coriano:2023hts, Coriano:2024ssu}.\\
As we move away from the conformal point, these interactions are characterized by specific sum rules satisfied by the corresponding anomaly form factors, which we examine in detail. Their spectral densities exhibit intriguing cancellation patterns between localized spectral poles and continuum contributions—structures that, to our knowledge, have not been previously analyzed with such precision. The spectral analysis is performed in full generality.  As we will demonstrate, these cancellations are closely tied to the structure of the effective action that is expected to reproduce the anomalies under broad kinematic conditions.

\subsection{Local and nonlocal effective actions}
A central question addressed in this work is whether anomaly poles appearing in the off-shell effective action—resembling axion-like interactions—can be associated with actual asymptotic axion-like states. This issue is particularly relevant given the strong interest in axion-like particles (ALPs) as potential dark matter candidates. The inclusion of such states on purely phenomenological grounds has often been proposed to explain various astrophysical observations (see the discussion in \cite{Coriano:2018uip}).\\
In standard quantum field theory, these particles are typically characterized by chiral anomaly interactions. In the conformal limit, such interactions can be analyzed using momentum-space conformal field theory ($CFT_p$), through 3-point functions governed by a nonlocal effective action. This action captures the essential features we investigate, including the emergence of anomaly poles and the associated sum rules.\\
A key part of our analysis is to demonstrate that the effective action is intrinsically nonlocal and does not, in general, support the existence of an asymptotic axion-like state. In particular, obtaining a local action from this structure would require enforcing conformal Ward identities that project out the transverse components of the anomaly vertex. However, doing so invalidates the sum rule central to our findings. The only viable path to a local effective action involves strong coupling effects that evade the UV constraints emerging from the conformal regime—mirroring the situation with standard Peccei–Quinn axions. In those cases, the modification of the functional measure (equivalent to including anomalous Ward identities) is crucial to deriving a consistent local action.\\
To clarify this further, we investigate the spectral densities of the anomaly form factors whose spectral densities manifest both anomaly poles and continuum contributions. Anomaly poles reflect the topological nature of anomaly vertices: they appear or vanish depending on the kinematic regime, governed by whether the relevant sum rule is saturated by a pole (discrete state) or by a continuum (dispersive cut). Our analysis shows that, in the perturbative regime, these interpolating states can be identified with asymptotic axion-like particles only when the interaction is constrained to the light-cone.
This unusual behavior stems from the intricate mixing between longitudinal and transverse sectors of the anomaly vertex, governed by Schouten identities. \\
These identities enable cancellations or enhancements in the residue structure of the form factors. Crucially, anomaly vertices are uniquely characterized by a sum rule always satisfied by their longitudinal form factors—a feature absent in purely local axion-like couplings of the form $a(x)\, F \tilde{F}$. While such local interactions relate to the full anomaly vertex via a Ward identity, they lack the spectral structure revealed by a dispersive analysis of both the longitudinal and transverse sectors. This distinction underlines the necessity of our nonlocal approach to fully capture the dynamics of anomaly-induced axion-like behavior.

\subsection{Anomaly Sum Rules}

Sum rules in the $AVV$ interaction have been previously studied in various kinematical configurations \cite{Dolgov:1971ri, Frishman:1980dq, Horejsi:1992tw}. A more general perspective on the chiral and conformal anomaly in the effective action is offered in \cite{Giannotti:2008cv}, with further investigations into the spectral density of anomaly form factors, including sum rules in the $\mathcal{N}=1$ superconformal anomaly multiplet and the Konishi anomaly \cite{Coriano:2014gja}.\\
Sum rules govern the absorptive amplitudes of anomaly vertices and manifest as area laws tied to the longitudinal form factors. In the $AVV$ vertex, we analyze the spectral densities of these form factors, identifying the structure of poles and cuts in the complex momentum plane ($q^2 \equiv t$), and their interplay before and after dispersion integration.
In this work, we build on these foundations by exploring different representations using Schouten identities, which can introduce or eliminate anomaly poles in vertex parameterizations. \\
We focus on their physical relevance, especially near the conformal limit, where anomaly poles are essential for satisfying conformal Ward identities \cite{Coriano:2023cvf, Coriano:2023hts, Coriano:2024ssu}. We adopt a minimal form factor basis and use $CFT_p$ methods for a sector decomposition into longitudinal and transverse components.\\
Notice that our approach leverages minimal decompositions \cite{Bzowski:2013sza}, optimized for the massless limit in parity-even trace-anomalous correlators, and recently extended to parity-odd sectors ($AVV$, $ATT$) via Schouten-based parameterizations \cite{Coriano:2023cvf, Coriano:2023hts, Coriano:2024ssu}. The transverse sectors are characterized by anomaly form factors, while the longitudinal ones encode the anomaly pole and satisfy the anomalous Ward identities. As we deviate from the conformal point, the dispersive analysis remains centered on these longitudinal sectors, where sum rules apply.
Finally, we present theoretical evidence supporting the interpretation of the interpolating states—described in \cite{Giannotti:2008cv} as involving pseudoscalar kinetic mixing—as effectively defined only on the light cone.

\section{Ordinary vs. Anomalous Couplings of Goldstone Modes}
In order to uncover the difference between ordinary (non anomalous) Goldstone modes and their coupling to the divergence of a (non anomalous) spontanously broken symmetry, and the axion case, 
we proceed with a comparison between the two cases. \\
In theories with spontaneously broken global symmetries and anomalies, Goldstone modes such as axions can couple to gauge fields through interactions like \( a(x) F \tilde{F} \). These couplings arise when a global chiral symmetry with a mixed anomaly—such as the axial-vector–vector $(AVV)$ anomaly—is spontaneously broken. The resulting Nambu–Goldstone (NG) boson couples to the anomaly through a mechanism rooted in Ward identities. This connection is often described using a local interaction, though the full anomaly structure in the currents currelator is fundamentally nonlocal.\\
 To illustrate this, consider a model with a fermion \( Q \) and a complex scalar field \( \Phi \), invariant under a global chiral \( U(1)_A \) symmetry that is anomalous. When \( \Phi \) acquires a vacuum expectation value, the symmetry is spontaneously broken, and the resulting phase degree of freedom \( \pi(x) \) behaves as a Goldstone boson. The fermion gains mass via a Yukawa interaction with \( \Phi \), while the chiral current \( J_A^\mu \) acquires both fermionic and bosonic components.\\
The Goldstone mode \( \pi(x) \) couples to the divergence of the axial current via
\beq
\mathcal{L} \sim \pi(x) \partial_\mu J_A^\mu.
\eeq
If \( J_A^\mu \) is anomalous, its divergence includes a term like \( F \tilde{F} \), and this gives rise to the axion–gauge field coupling
\beq
\mathcal{L} \sim \frac{a(x)}{v} F \tilde{F}.
\eeq

However, this derivation simplifies the full structure of the anomaly. In the perturbative picture, the anomaly is encoded in the longitudinal part of a three-point function (such as the \( AVV \) vertex), which contains a \(1/q^2\) pole indicating the exchange of a massless pseudoscalar. This is not a true particle pole but rather a signal of the nonlocal nature of the anomalous interaction. Its presence arises from solving the anomalous Ward identity
\beq
q_\lambda \Delta^{\lambda \mu \nu}(q, p_1, p_2) \sim a_1 \epsilon^{\mu \nu p_1 p_2},
\eeq
which admits a solution of the form
\beq
\Delta^{\lambda \mu \nu} \sim \frac{q^\lambda}{q^2} \epsilon^{\mu \nu p_1 p_2}.
\eeq
When mass terms are included, the divergence of the axial current gains a contribution from the pseudoscalar density
\beq
\partial_\mu J_A^\mu = a_1 F \tilde{F} + 2 m_Q \bar{Q} \gamma_5 Q,
\eeq
that modifies the equation of motion for the Goldstone boson, by adding a mass correction via coupling to the pseudoscalar current. Consequently, the effective interaction becomes
\beq
\mathcal{L} \sim \frac{a(x)}{v} \left( F \tilde{F} + 2 m_Q \bar{Q} \gamma_5 Q \right).
\eeq

In momentum space, this is reflected in a vertex structure involving both the massless anomaly term and a massive correction encoded in the scalar loop integral \( C_0 \) of a three-point function (see \eqref{lco}). Indeed, the full axion–photon–photon vertex becomes
\beq
V^{\mu\nu} \sim \epsilon^{\mu \nu p_1 p_2} \left( 1 + 2 m^2 C_0(p_1^2, p_2^2, q^2, m^2) \right).
\eeq
While the first term corresponds to the anomaly, the second arises from explicit chiral symmetry breaking via mass terms.
Importantly, the anomaly pole (i.e., the \(1/q^2\) behavior) dominates in the conformal or massless limit. When fermion masses or off-shell momenta are included, the residue of the pole typically vanishes. This implies that the anomaly pole is not a true propagating particle, but it still defines the dominant asymptotic behavior of the vertex at large momentum transfer.\\
Furthermore, the local interaction \( a(x) F \tilde{F} \), while useful phenomenologically (e.g., in axion models like Peccei–Quinn), does not fully capture the rich tensor structure of the anomaly, especially the interplay between longitudinal and transverse components governed by conformal symmetry. The complete off-shell correlator contains both, and they are linked by the anomaly coefficient in a way that the local description obscures.\\
In contrast to NG modes from non-anomalous symmetries - which couple derivatively and decouple in the infrared - anomalous NG modes retain a distinct signature due to the anomaly pole. However, as we have mentioned above, this pole does not define a physical asymptotic state unless specific kinematic conditions are met that require a light-cone dynamics. Away from these limits, the full vertex must be retained to describe the interaction accurately.\\
Finally, while the use of Ward identities to derive local interactions is consistent at tree level or in effective field theory, the full quantum structure of anomaly vertices—especially in gravitational or mixed gauge-gravitational contexts—requires a nonlocal formulation. This ensures correct matching to perturbative computations, respects the topological origin of the anomaly, and captures sum rules and dispersion relations not visible in local couplings.

\section{Reconstructing the \(AVV\) Correlator from the Anomaly Pole and Conformal Symmetry: review}

We begin our analysis in the conformal limit, where the conformal constraints in momentum space, combined with the inclusion of a pole in the solution of these identities for the \(AVV\) correlator, play a crucial role.  
In this regime, the conformal constraints enable the complete identification of parity-odd anomaly interactions using only a single input—an anomaly pole in the longitudinal sector of each correlator. This result is inherently nonperturbative, as it does not rely on any specific Lagrangian formulation. However, once we move away from the conformal point and introduce massive fermions in virtual corrections, this picture undergoes significant modifications.  
Before delving into these corrections, which will allow for a deeper exploration of the sum rule associated with this correlator, we first provide a brief review of how the presence of a pole, together with the conformal Ward identities, enables the reconstruction of anomaly interactions in the conformal limit. This approach further facilitates the characterization of all form factors in the decomposition of these correlators into longitudinal and transverse sectors. \\
We start by imposing the external Ward identities in flat space-time for a vector and an axial-vector current respectively as quantum averages
\begin{align}\label{eq:CWIJJJ}
\partial_\mu \braket{J^\mu}=0, \qquad\qquad \partial_\mu \braket{J^\mu_A}=a_1 \, \varepsilon^{\mu\nu\rho\sigma}F_{\mu\nu}F_{\rho\sigma}.
\end{align}
 The Ward identities for the $AVV$ are extracted from \eqref{eq:CWIJJJ} by functional differentiation 
\begin{equation}\label{eq:jjjconsward3p}
\begin{aligned}
&p_{i\mu_i}\,\braket{J^{\mu_1}(p_1)J^{\mu_2}(p_2)J_A^{\mu_3}(p_3)}=0,\quad \quad i=1,2\\[1.2ex]
&p_{3\mu_3}\,\braket{J^{\mu_1}(p_1)J^{\mu_2}(p_2)J_A^{\mu_3}(p_3)}=-8i  \, a_1  \, \varepsilon^{p_1p_2\mu_1\mu_2},
\end{aligned}
\end{equation}
where $a_1$ is the anomaly coefficient. \\
Whenever we will be using a symmetric $(1,2,3)$ labeling of the momenta, useful for the conformal parameterization of the vertices, all the momenta $p_i$ will be chosen as incoming.  \\ 
In order to consider also a combined gauge and gravitational background, as in  the case for the gravitational anomaly, the quantum average of the axial vector current will be extended with the inclusion of a covariant derivative and 
curvature related terms 
\begin{equation} \label{eq:anomaliachirale}
	\nabla_\mu\langle J_A^\mu\rangle=
	a_1\, \varepsilon^{\mu \nu \rho \sigma}F_{\mu\nu}F_{\rho\sigma}+ a_2 \, \varepsilon^{\mu \nu \rho \sigma}R^{\alpha\beta}_{\hspace{0.3cm} \mu \nu} R_{\alpha\beta \rho \sigma},
\end{equation}
that will be investigated when we will deal with the $ATT$ correlator. 
As already mentioned, in both the case of the $ATT$ and $AVV$, these conditions, together with the ordinary Ward identities imposed either from gauge invariance or diffeomorphism invariance, set the boundary conditions for the 
reconstruction of the corresponding correlators by the inclusion of a single anomaly pole. \\ 
In the $AVV$ case, from these relations we construct the general form of the correlator, splitting the operators into 
a transverse and a longitudinal part as
\begin{equation}
\begin{aligned}
\label{ex}
&J^{\mu}(p)=j^\mu(p)+j_{loc}^\mu(p),\\
&j^{\mu}=\pi^{\mu}_{\alpha}(p)\,J^{\alpha}(p),\quad \pi^{\mu}_\alpha(p)\equiv \delta^{\mu}_\alpha-\frac{p_\alpha\,p^\mu}{p^2},\\
&j_{loc}^{\mu}(p)=\frac{p^\mu}{p^2}\,p\cdot J(p),
\end{aligned}
\end{equation}
and
\begin{equation}
\begin{aligned}
\label{ex}
&J_A^{\mu}(p)=j_A^\mu(p)+j_{A \,\text{loc}}^\mu(p),\\
&j_A^{\mu}=\pi^{\mu}_{\alpha}(p)\,J_A^{\alpha}(p),\\
&j_{ A \,\text{loc}}^{\mu}(p)=\frac{p^\mu}{p^2}\,p\cdot J_A(p).
\end{aligned}
\end{equation}
Due to \eqref{eq:jjjconsward3p}, the correlator is purely transverse in the vector currents. We then have the following decomposition
\begin{equation}
	\braket{ J^{\mu_1 }(p_1) J^{\mu_2 } (p_2)J_A^{\mu_3}(p_3)}=\braket{ j^{\mu_1 }(p_1) j^{\mu_2 }(p_2) j_A^{\mu_3}(p_3)}+\braket{J^{\mu_1 }(p_1) J^{\mu_2 }(p_2)\, j_{A \text { loc }}^{\mu_3}(p_3)}\label{decomp}
\end{equation}
where the first term is completely transverse with respect to the momenta $p_{i\mu_i}$, $i=1,2,3$ and the second term is the longitudinal part that is proper of the anomaly contribution.
 Using the anomaly constraint \eqref{eq:jjjconsward3p} on $j_{A loc}$ we obtain the expression of the longitudinal component of 
 \eqref{decomp}
\begin{align}
\label{anomc}
\braket{J^{\mu_1 }(p_1) J^{\mu_2 }(p_2)\, j_{A \text { loc }}^{\mu_3}(p_3)}=\frac{p_3^{\mu_3}}{p_3^2}\,p_{3\,\alpha_3}\,\braket{J^{\mu_1}(p_1)J^{\mu_2}(p_2)J_A^{\alpha_3}(p_3)}=\Phi_0\,\varepsilon^{p_1p_2\mu_1\mu_2}\,p_3^{\mu_3},
\end{align}
with 
\beq
\Phi_0=-\frac{8i\,a_1}{p_3^2}
\eeq
defining the anomaly form factor in the conformal limit. The complete solution can be obtained by solving the special conformal Ward identities and using the Schouten relations in order to parameterize also the transverse sector of  
\eqref{decomp}. The procedure involves an intermediate step in which the action of the special conformal transformation on the correlator is projected onto the transverse sector and separated into second order and first order equations using the Schouten relations.Then, the general structure of the transverse part can be further simplified using the Schouten identities, resulting in the expression \cite{Coriano:2023hts}

 \begin{align}
\label{ref}
	\braket{J^{\mu_1}(p_1)J^{\mu_2}(p_2) J^{\mu_3}_A (p_3)} &=-\frac{8i\,a_1}{p_3^2}\varepsilon^{p_1p_2\mu_1\mu_2}\,p_3^{\mu_3} \nonumber \\
	&8 i a_1\,\pi^{\mu_1}_{\alpha_1}(p_1)
	\pi^{\mu_2}_{\alpha_2} (p_2) \pi^{\mu_3}_{\alpha_3}
	\left(p_3\right)\bigg[ \,p_2^2\,I_{3\{1,0,1\}}\, \varepsilon^{p_1 \alpha_1\alpha_2\alpha_3}-
	\,p_1^2\,I_{3\{0,1,1\}}\, \varepsilon^{p_2\alpha_1\alpha_2\alpha_3}  
	\bigg], \nonumber \\
\end{align}
where, remarkably, the $a_1$ the anomaly coefficient has turned into a structural factor both for the longitudinal and transverse parts.
Unlike the perturbative approach, where the anomaly coefficient is primarily determined by the prefactor in front of a Feynman amplitude, here it emerges as the residue of a pole in the longitudinal sector, as a solution of the anomaly constraint. The conformal Ward identities then propagate this pole to the other sectors of the correlator, specifically the transverse one.  \\
Notably, this parameterization of the \(AVV\) correlator is the most compact one, as it not only highlights the central role of the anomaly pole but also reveals additional features that we will discuss shortly.
The solution involves the integral
 \begin{align}
&I_{3\left\{1,0,1\right\}}=p_1\,p_3\frac{\partial^2}{\partial p_1\partial p_3}\,I_{1\{0,0,0\}},
\end{align}
where 
$I_{1\{0,0,0\}}$ is the master integral (3K integral) related to the three-point function in the theory of a free massless scalar field in $d=4$ \cite{Bzowski:2013sza}
\begin{align}
	I_{1\{0,0,0\}}=\frac{1}{4}C_{0}(p_1^2,p_2^2,p_3^2).
\end{align}
One can then express the 3K integral explicitly as follows
\begin{align}
	I_{3\left\{1,0,1\right\}}(p_1^2,p_2^2,p_3^2)=&\frac{1}{\lambda^2}\bigg\{-2 p_1^2 p_3^2 \bigg[p_1^2 \left(p_2^2-2 p_3^2\right)+p_1^4+p_2^2 p_3^2-2 p_2^4+p_3^4\bigg] C_0\left(p_1^2,p_2^2,p_3^2\right)\notag\\
	&+p_1^2 \left(\left(p_1^2-p_2^2\right)^2+4p_2^2p_3^2- p_3^4\right)\log\left(\frac{p_1^2}{p_2^2}\right)+4p_1^2 p_3^2 \left(p_1^2-p_3^2\right)\log\left(\frac{p_1^2}{p_3^2}\right)\notag\\
	&-p_3^2 \Big((p_2^2-p_3^2)^2+4p_1^2p_2^2-p_1^4\Big)\log\left(\frac{p_2^2}{p_3^2}\right)-\lambda(p_1^2-p_2^2+p_3^2)\bigg\}
\end{align}
where
\begin{equation}
\label{lco}
		C_0\left(q^2,p_1^2,p_2^2\right)= \frac{1}{i \pi^2}\int d^n k\frac{1}{k^2 (k- q)^2 (k - p_1)^2},\,
	\end{equation}
 is the the ordinary massless scalar master integral. It is given by
\beq
C_0(s,s_1,s_2) = \frac{1}{s} \Phi(x,y).
\eeq
$\Phi(x,y)$ is the special function  \cite{Usyukina:1992jd} 
\bea
\Phi( x, y) &=& \frac{1}{\lambda} \biggl\{ 2 [\textrm{Li}_2(-\rho  x) + \textrm{Li}_2(- \rho y)]  +
\ln \frac{y}{ x}\ln \frac{1+ \rho y }{1 + \rho x}+ \ln (\rho x) \ln (\rho  y) + \frac{\pi^2}{3} \biggr\},
\label{Phi}
\eea
where
\bea
 \lambda(1,x,y) = \sqrt {\Delta},
 \qquad  \qquad \Delta=(1-  x- y)^2 - 4  x  y,
\label{lambda} \\
\rho( x,y) = 2 (1-  x-  y+\lambda)^{-1},
  \qquad  \qquad x=\frac{s_1}{s} \, ,\qquad \qquad y= \frac {s_2}{s}\qquad s\equiv p_3^2\, .
\eea
The analysis highlights the pivotal role of the anomaly pole in determining the entire anomaly interaction, particularly in the conformal limit. Remarkably, the complete off-shell correlator is fully specified by a single additional form factor, which is connected to the integral \( I_{3\{1,0,1\}} \).
Notably, the conformal solution is inherently scale-independent and entirely dictated by the anomaly coefficient. If this coefficient is set to zero, the correlator vanishes entirely.\\
At this juncture, a crucial question arises: does this $1/p_3^2\equiv 1/q^2$ interaction correspond to an asymptotic state?\\
 For such an interpretation, the anomaly pole must be identified as a massless particle pole. To establish this, one must verify under general kinematic conditions whether the residue of the vertex 
\beq
\label{gpp}
g_{p} \equiv\lim_{q^2\to 0} q ^2 \braket{J^{\mu_1}(p_1)J^{\mu_2}(p_2) J^{\mu_3}_A (q)} \neq 0
\eeq
is nonvanishing.
The subtle nature of the anomaly interaction lies in the fact that the longitudinal form factor can exhibit an anomaly pole, even though \eqref{gpp} may vanish. As already pointed out in the previous sections, anomaly poles do not necessarily correspond to particle poles. As a result, the on-shell limits used to identify the anomaly coupling in the 1PI vertex are deeply connected to the properties of this vertex in its complete tensorial form rather than just with its longitudinal part. 
\\
To move away from the conformal limit, we address the massive fermion case and compute the vertex using ordinary perturbation theory. 
We can perform a decomposition similar to \eqref{ref} in the form 
\begin{align}
\label{refm}
	\braket{J^{\mu_1}(p_1)J^{\mu_2}(p_2) J^{\mu_3}_A (p_3)} &=	\braket{J^{\mu_1 }(p_1) J^{\mu_2 }(p_2)\, j_{A \text { loc }}^{\mu_3}(p_3)}
 	+ \pi^{\mu_1}_{\alpha_1}(p_1)\pi^{\mu_2}_{\alpha_2} (p_2) \pi^{\mu_3}_{\alpha_3}(p_3)
	\, \Delta_T^{\alpha_1 \alpha_2 \alpha_3 }(p_1^2,p_2^2,p_3^2,m^2) 
	 \nonumber \\
\end{align}
with the longitudinal sector given by 
\begin{equation}
\label{nonloc}
		\braket{J^{\mu_1 }(p_1) J^{\mu_2 }(p_2)\, j_{A \text { loc }}^{\mu_3}(p_3)}\equiv \Phi_0(p_1^2,p_2^2,p_3^2,m^2) \varepsilon^{p_1p_2\mu_1\mu_2}\,p_3^{\mu_3}.
	\end{equation}
  The expression of the transverse part is denoted by $\Delta_T$.\\
 To go beyond the conformal limit, we return to perturbation theory by introducing a fermion of mass \( m \). The \( AVV \) anomaly tensor \(\Delta_{\alpha\mu\nu}(p_1, p_2)\) is defined through the triangle fermion loop
\begin{equation}
\label{orig}
\Delta_{\alpha\mu\nu}(p_1, p_2) = -\int \frac{d^4 k}{(2\pi)^4} \mathrm{Tr} \left( \frac{1}{\not{k} - \not{p_1} - m} \gamma_\mu \frac{1}{\not{k} - m} \gamma_\nu \frac{1}{\not{k} + \not{p_2} - m} \gamma_\alpha \gamma_5 \right) + [ (p_1, \mu) \leftrightarrow (p_2, \nu) ].
\end{equation}

The vector current remains conserved
\beq
p_1^\mu \Delta_{\alpha\mu\nu} = 0, \quad p_2^\nu \Delta_{\alpha\mu\nu} = 0,
\eeq
while the axial-vector Ward identity is anomalous
\beq
q^\alpha \Delta_{\alpha\mu\nu} = 2m \Delta_{\mu\nu}(p_1, p_2) + \frac{1}{2\pi^2} \epsilon_{\mu\nu\rho\sigma} p_1^\rho p_2^\sigma.
\eeq

The tensor structure of \(\Delta_{\alpha\mu\nu}\) is parameterized as
\begin{equation}
\label{deff}
\Delta_{\alpha\mu\nu} = F_1(s, m^2)\, q_\alpha \epsilon_{\mu\nu\rho\sigma} p_1^\rho p_2^\sigma 
+ F_2(s, m^2)\, (p_{2\nu} \epsilon_{\alpha\mu\rho\sigma} - p_{1\mu} \epsilon_{\alpha\nu\rho\sigma}) p_1^\rho p_2^\sigma.
\end{equation}

The anomaly resides entirely in the \( F_1 \) form factor, which obeys
\begin{equation}
\label{broken}
s F_1(s, m^2) = -\frac{1}{\pi^2} \int_0^1 dx \int_0^{1-x} dy \, \frac{m^2}{m^2 - x y s - i\epsilon} + \frac{1}{2\pi^2}.
\end{equation}

The second term on the rhs of the equation above defines the anomaly, while the first accounts for explicit chiral symmetry breaking. In the massless limit, only the anomaly term survives and saturates the anomaly sum rule.

The off-shell generalization of \( F_1 \), denoted as \( \Phi_0 \), is
\begin{equation}
\label{fst}
\Phi_0(p_1^2, p_2^2, p_3^2, m^2) = \frac{m^2}{\pi^2} \frac{1}{q^2} C_0(q^2, p_1^2, p_2^2, m^2) + \frac{1}{2\pi^2} \frac{1}{q^2},
\end{equation}
with \( C_0 \) the standard scalar triangle integral \eqref{lco} parameterized as
\begin{equation}
C_0(q^2, p_1^2, p_2^2, m^2) = \int_0^1 dx \int_0^{1-x} dy \, \frac{1}{\mathcal{D}(x,y)},
\end{equation}
where
\beq
{D}(x,y) = m^2 + q^2 y(x + y - 1) - p_1^2 x y + p_2^2 x(x + y - 1).
\eeq

While \eqref{fst} exhibits a \(1/q^2\) term resembling an anomaly pole, this feature is misleading, since the form factor is characterized by a branch cut rather than a true pole in the massive theory. The full cancellation of this apparent pole will be explored later, and is closely related to the anomaly sum rule. For comparison, the local form factor derived from the axial-vector Ward identity is
\begin{align}
\label{loc}
q_\lambda \braket{J^\mu J^\nu J_A^\lambda} &\equiv \Phi_W\, \epsilon^{\mu\nu p_1 p_2}, \\
\Phi_W &= q^2 \Phi_0 = \frac{m^2}{\pi^2} C_0(p_1^2, p_2^2, q^2, m^2) + \frac{1}{2\pi^2}.
\end{align}
Unlike \( \Phi_0 \), the local form \( \Phi_W \) contains no \(1/q^2\) structure and does not obey an anomaly sum rule. It is commonly used in effective axion–gauge field interactions, but lacks the nonlocal and topological properties intrinsic to \( \Phi_0 \).
Thus, while \( \Phi_W \) reflects a local vertex structure, only \( \Phi_0 \) captures the full anomaly content, including its asymptotic \(1/q^2\) behavior and relation to the anomaly pole. The distinction between these two forms is central to understanding the anomaly both in conformal and massive regimes.

\subsection{Pole cancellation in $\Phi_0$ for $m\neq 0$ for on-shell vectors}
We now prove that there is no pole in $\Phi_0$ when the two vector lines are on-shell and $m\neq 0$.
One can 
show that there is a cancelation between the second and the first contribution in $\Phi_0$ in 
\eqref{fst}.
For this purpose we consider the on shell case, $p_1^2=p_2^2=0$ and use the explicit expression of $C_0$ in the form 
\beq
\label{exp0}
C_0(0,0,q^2,m^2)\equiv C_0(q^2,m^2)=-\frac{1}{2 m^2}\,  {_3F_2} \left( 1, 1, 1; 2, \frac{3}{2}; x \right), 
\eeq
where the \(_3F_2\) hypergeometric function can be written as a series expansion for
 \( q^2/4m^2 < 1\) as 
\beq
_3F_2 \left( 1, 1, 1; 2, \frac{3}{2}; x \right) = \sum_{k=0}^\infty \frac{(1)_k (1)_k (1)_k}{(2)_k \left(\frac{3}{2}\right)_k} \frac{x^k}{k!},
\eeq
with \(x = \frac{p_2^2}{4m^2}\), and where \((a)_k\) is the Pochhammer symbol, defined as
\beq
(a)_k = a(a+1)(a+2)\cdots(a+k-1), \quad (a)_0 = 1.
\eeq
Substituting the Pochhammer symbols for the given parameters
\beq
\begin{aligned}
(1)_k &= 1 \cdot 2 \cdot 3 \cdots (k) = k!, \\
(2)_k &= 2 \cdot 3 \cdot 4 \cdots (k+1) = (k+1)k!, \\
\left(\frac{3}{2}\right)_k &= \frac{3}{2} \cdot \frac{5}{2} \cdot \frac{7}{2} \cdots \left(\frac{3}{2} + k - 1\right).
\end{aligned}
\eeq
one derives the expansion 
\beq
\label{exp1}
C_0(q^2, m^2) = -\frac{1}{2m^2} \left( 1 + \frac{q^2}{12m^2} + \frac{q^4}{180m^4} + \mathcal{O}\left(\frac{q^6}{m^6}\right) \right),
\eeq
giving an anomaly form factor of the form
\beq
\Phi_0 = -\frac{1}{4\pi^2} \frac{1}{12m^2} - \frac{1}{4\pi^2} \frac{q^2}{180m^4} +\ldots
\eeq
Notice that there has been a cancelation between the $1/q^2$ term in \eqref{fst} and a similar term coming from the scalar loop $C_0$. In other words, the 1PI effective interaction is entirely mass dependent. Given the condition 
$q^2< 4 m^2$, the expansion is valid below the cut. Notice that the form factor vanishes as the fermion mass is sent to infinity, indicating the decoupling of the heavy fermion in the perturbative loop.\\ 
The area subtended by the spectral density of $\Phi_0$ (see Fig 1) over the cut plays a crucial role in the analysis of the sum rule. Therefore, we now turn to a more general examination of this form factor.
\begin{figure}[t]
\centering
\subfigure[]{\includegraphics[scale=0.8]{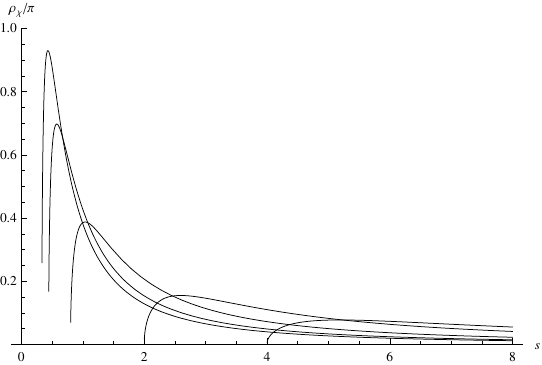}}  \hspace{2cm}
\caption{Spectral density flow in the $p_1^2 = p_2^2 = 0$ case as a function of the fermion mass $m$. The sum rule is an area law for this spectral flow as one moves towards the conformal point  $m\to 0$.}
\label{dd1}
\end{figure}
\noindent

 \section{Chiral sum rules }
 In this section, we begin our investigation of the sum rules governing the anomaly form factors, starting with the case of the $AVV$ correlator. 
 The original derivation of the sum rule for the $AVV$ had been discussed for some specific 
 cases. Horejsi  \cite{Horejsi:1985qu} investigated the case $p_1^2=p_2^2=0, m\neq 0$. Previous analysis had derived the sum rule when only a single vector line is off-shell $(p_1^2=0,p_2^2\neq 0, m\neq 0)$ \cite{Dolgov:1971ri} and for 
 $(p_1^2=p_2^2 <0, m=0)$ \cite{Frishman:1980dq}. The area law implicit in this specific feature of the anomaly interactions is, at least in part, a result of the property of analiticity of the vertex but also manifests a correlation between the longitudinal and transverse sectors of the interaction. This point has been observed in \cite{Coriano:2014gja}. 
\\
The analiticity conditions of $\Phi_0$ in the complex $q^2$ plane is summarised by
the following two formulas
 \beqa
\oint_C \Phi_0(s,s_1,s_2,m^2)  \, ds & =& 0,
 \eeqa
and
 \beqa
 \label{disp}
 \Phi_0(q^2,s_1,s_2,m^2) &=& \frac{1}{2\pi i} 
    \oint_C \frac{ \Phi_0(s,s_1,s_2,m^2)}{s - q^2} \, ds, \nn\\
\eeqa
where the contour of integration, by Jordan's lemma, is reduced to the $s\geq 0$ axis, where the function, as we are going to show, exhibits poles and a cut for $s> 4 m^2$.  
As we are going to see, the spectral density 
 will contain poles in its rational factors, with a branch cut starting at $4 m^2$.

\begin{figure}
\label{figg1}
	\centering
	\begin{tikzpicture}
	
	\draw[thick, ->] (-1,0) -- (6,0) node[right] {$\text{Re}(s)$};
	\draw[thick, ->] (0,-3) -- (0,3) node[above] {$\text{Im}(s)$};
	
	\filldraw[red] (0,0) circle (3pt) node[below left] {$0$};
	
	\draw[thick, decorate, decoration={zigzag, amplitude=2pt, segment length=4pt}] (2,0) -- (5.9,0);
	\node at (2, -0.3) {$4m^2$};
	
	\draw[very thick, ->, red] (0,0) -- (0,1.5);     
	\draw[very thick, ->, red] (0,0) -- (0,-1.5); 
		\end{tikzpicture}
	\caption{Analyticity region of the on-shell anomaly form factor in the complex \( s\equiv q^2 \)-plane, with two bold spikes at \(s=0\) of different colour, for visual clarity. The sum rule of the anomaly form factor of the $AVV$ involves a cancelation between the two red spikes at $s=0$, leaving only the integral over the continuum for $s> 4 m^2$.} 
\end{figure}
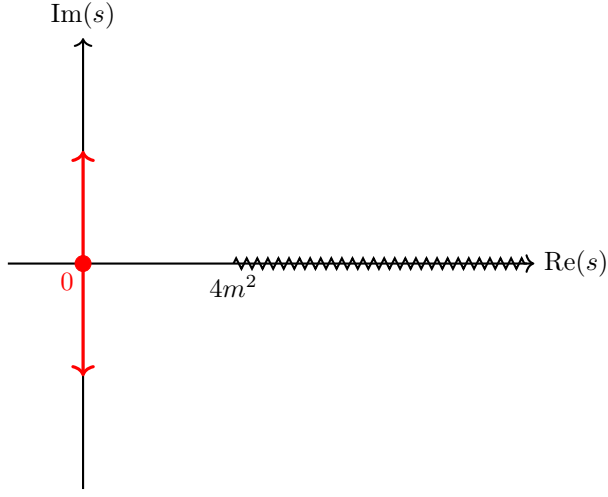
We start by defining the general discontinuity of $\Phi_0$ for $q^2 >0$, as
\beqa
  \textrm{disc}\Phi_0 
 &=& \Phi_0(q^2 + i \epsilon, p_1^2,p_2^2, m^2)- \Phi_0(q^2 - i \epsilon, p_1^2,p_2^2, m^2) 
 \eeqa
 and its imaginary part, the spectral density  $\Delta$, in the $s$ variable as
 \beq
 \Delta \Phi_0(s,s_1,s_2,m^2)\equiv \Im \Phi_0(s,s_1,s_2,m^2) = \frac{1}{2 i}\textrm{disc}\,\Phi_0.
 \eeq
 The vanishing of the boundary integral on the large radius allows to rewrite the dispersion relation \eqref{disp} as

 \beq
 \Phi_0(q^2,p_1^2,p_2^2,m^2)=\frac{1}{\pi}\int_0^\infty \frac{\Delta\Phi_0(s, p_1^2,p_2^2,m^2)}{s - q^2}{ds}.
 \eeq
 As previously mentioned, the relation holds in the complex $q^2$ plane at fixed positive values of $s_1$ and $s_2$. Notice that the dispersion relations we derive are unsubtracted, as we enforce vector Ward identities to re-express the divergent form factors in terms of convergent ones. The computation of the spectral density of the anomaly form factors is carried out using the principal value prescription for the $1/q^2$ pole appearing in their explicit expressions
 \beq
 \label{ddd}
 \textrm{disc}\, \frac{1}{q^2}\equiv \frac{1}{q^2 + i \epsilon}- \frac{1}{q^2  - i \epsilon}=- 2 \pi i \delta(q^2).
 \eeq
 Using this relation in \eqref{fst} 
  \beqa
 \label{zero}
 \textrm{disc}\,\Phi_0 (q^2,p_1^2,p_2^2, m^2)
 &=& \frac{m^2}{\pi^2} \textrm{disc}\left( \frac{C_0(q^2,p_1^2,p_2^2,m^2)}{q^2}\right) + \frac{1}{2 \pi^2}\textrm{disc}\frac{1}{q^2}
 \eeqa
 with the help of \eqref{ddd}, we can separate the pole and the continuum contributions  as 
 \beq
 \label{first}
 \textrm{disc}\,\Phi_0(q^2,p_1^2,p_2^2, m^2)= - 2 i \frac{ m^2}{\pi} \, C_0 (q^2,p_1^2,p_2^2, m^2)\, \delta(q^2)  -  \frac{i}{\pi}\, \delta(q^2)
+\frac{ m^2}{ \pi^2 q^2}\textrm{disc} \,C_0(q^2,p_1^2,p_2^2, m^2),
 \eeq
showing that the spectral density exhibits two distinct localized terms at $s=0$ ($\delta$ spikes) and 
a continuum contribution for $q^2 > 4 m^2$, with an interplay among these three components, as we are going to show, which is related to the property of analiticity of $\Phi_0$. We will refer to the localized components as
\beq
\label{spk}
\begin{aligned}
    \text{spk}_1 &\equiv - \frac{i}{\pi} \, \delta(q^2), \\
  \text{spk}_2 &\equiv - 2 i\frac{ m^2}{\pi} \, C_0(q^2, p_1^2, p_2^2, m^2) \, \delta(q^2). 
\end{aligned}
\eeq
The first spike is of constant strength, while the second plays a role in equaling the integral of the density over the cut, causing a cancelation. One can think of these two components as distinct but superimposed resonances at $s=0$.
These features, as we shall demonstrate, either before or after integration, exhibit cancellation patterns tied to the external kinematics of the vertex. Such cancellations arise either within the two spikes themselves or through their interaction of  $\text{spk}_2$ with the continuum. 
The continuum corresponds to the discontinuity in the scalar loop, \(\textrm{disc} \, C_0\), which is determined by placing two intermediate lines of the triangle diagram on-shell.

\subsection{The use of the Laplace/Borel transforms }
\label{lap}
 The on-shell behaviour presented above can be discussed for arbitrary kinematics. The pattern of cancelations illustrated in 
 Fig. 2 takes a different form in the general case, when the vector lines are off-shell and we are away from the conformal point. 
 In this case the cancelation is between the integral of $\textrm{spk}_2$ and the one over the continuum, and is illustrated in Fig. 3.  
 To prove this point we introduced a method that allows to establish this result starting directly from the parametric integral. 
 This method is essentially based on the use of Borel transforms.\\
  The evaluation of the integral of the spectral density of $\Phi_0$ along the 
$\left[ 4 m^2,\infty)\right. $ cut
can be computed by a trick that we are going to describe, using the inverse Laplace transform of the same density (indicated by $t$), followed  by the $t\to 0$  limit on the same variable. For this, in full generality, we define the inverse Laplace transform $(f(t))$ of a generic function 
$F(s)$

\begin{equation}
			f(t) = \mathcal{L}^{-1}\{F(s)\} = \frac{1}{2\pi i} \int_{\gamma - i \infty}^{\gamma + i \infty} F(s) e^{st} \, ds,
	\end{equation}
and recall that for a rational function $F(s)$ with a pole at $s={s_0}$ 	
	\begin{equation}
		\mathcal{L}^{-1}\{\frac{1}{s-{s_0}}\}(t)=e^{{s_0}t}
	\end{equation}
	which provides a nonzero contribution to the dispersion integral as $t\to 0$. 
The method allows to compute the integral over the continuum of the spectral density quite efficiently, but it requires some care since extra singular contributions, that can be viewed as subtractions in the reconstruction of a given form factor, are not identified by this approach. For example, double poles of the form $1/(s-{s_0})^2$ are not taken into account by the method since one obtains
\begin{equation}
		\mathcal{L}^{-1}\{\frac{1}{(s-{s_0})^2}\}(t)=t e^{{s_0}t},
	\end{equation}
that vanishes in the same $t\to 0$ limit.  An example can be found in the analysis of the $ATT$ case, where 
we discuss how this point can be handled (\secref{trash}). \\
 In our case, for the anomaly form factor $\Phi_0$, given a spectral representation of the form
\beq
\Phi_0(q^2)=\frac{1}{\pi}\int_0^{\infty} \frac{\Delta \Phi_0(s)}{s- q^2}  \,ds
\eeq
the action of the inverse transform will generate the expression
	\begin{equation}
		\mathcal{L}^{-1}\Bigl\{  \frac{1}{\pi} \int_0^{\infty} \frac{\Delta \Phi_0}{s-q^2}  \,ds \Bigl\}(t)  = \frac{1}{\pi} \int_0^{\infty} 
		\Delta \Phi_0(s) e^{s t}  \,ds	\end{equation}
and the integral of the spectral density will be obtained by a limit on $t$
\beq	
\lim_{t\rightarrow 0}\mathcal{L}^{-1}\{\Phi_0\}(t)  =\frac{1}{\pi}\int_0^{\infty} \Delta \Phi_0  \,ds 
\eeq
This approach allows to derive the expression of the integral of the discontinuity along the cut and verify the sum rules of such form factors. \\
For this purpose we use the parametric representation 
\begin{equation}
	\label{parm}
			C_0\left(p_1^2,p_2^2,q^2,m^2,m^2,m^2\right)= \int_0^1 dx\, \int _0^{1-x} dy\,  \frac{1}{\D(x,y)}
	\end{equation}
	
	\begin{equation}
		\D(x,y)=m^2+q^2 y (x+y-1)-p_1^2 x y+p_2^2 x (x+y-1)
	\end{equation}

that we insert in the expression of $\Phi_0$ given in \eqref{fst}. A direct computation gives 
for the inverse Laplace transform the expression
	\begin{equation}
		\mathcal{L}^{-1}\{\Phi_0\}(t)=-\int_0^1 dx \int_{0}^{1-x} d y \, \frac{g^3 \left( m^2 \exp \left(\frac{t \left(-m^2+p_1^2x y-p_2^2 x^2-p_2^2  x y+p_2^2  x\right)}{x y+y^2-y}\right)-2 m^2+p_1^2 x y-p_2^2  x^2-p_2^2  x y+p_2^2 x\right)}{\pi ^2 \left(m^2-p_1^2 x y+p_2^2  x^2+p_2^2 x y-p_2^2  x\right)}
	\end{equation}
on which we perform the $t\to 0$ limit as described above, in order to derive directly the value of the integral of the discontinuity over the entire cut	
		\begin{equation}
		\label{sumr1}
	\mathcal{L}^{-1}\{\Phi_0\}(t\to 0)=\frac{1}{\pi}\int_{4 m^2}^\infty  \Delta{\Phi_0}(s, p_1^2,p_2^2,m^2) ds  =\frac{1}{2\p^2}.
	\end{equation}
Eq. \eqref{sumr1} holds in the most general case, for off-shell vector lines and a massive fermion. This result is in agreement with the analysis of \cite{Giannotti:2008cv}.\\	
 Notice that a sum rule holds only for $\Phi_0$ and not for $\Phi_W$, the form factor extracted from the local formulation of the anomaly interaction \eqref{loc}.
A direct computation indeed shows that the corresponding parametric integral, in this case,
\beq
\int_{4 m^2}^\infty \Delta{\Phi_W}(s, p_1^2,p_2^2,m^2) ds=
 \int_0^1 \left(\int_0^{1-x} \frac{g^3 m^2}{2 \pi ^2 \left(x y+y^2-y\right)} \, dy\right) \, dx 
\eeq
is indeed divergent.  \\
\begin{figure}[t]
\centering
\subfigure[]{\includegraphics[scale=0.8]{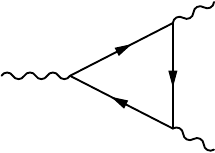}}  \hspace{2cm}
\subfigure[]{\includegraphics[scale=0.8]{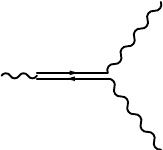}} \hspace{2cm}
\subfigure[]{\includegraphics[scale=0.8]{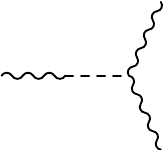}}
\caption{The fermion loop (a); the collinear region (b); the effective pseudoscalar exchange (c). The $q^2=0$ point defining the residue of the interaction can be viewed as a massless state decaying into a massless virtual pseudoscalar state that couples to the anomaly at the last stage.}
\label{ddiag}
\end{figure}

\section{The residue of the particle pole in the light-cone variables }  
The computation of the residue of a possible particle pole in the off-shell massive $AVV$ correlator requires the choice of special noncovariant frames where to proceed with the analysis. This allows to test if the coupling 
$g_{p}$ in \eqref{gpp} is nonzero. The computation of the residue of the particle pole of the $AVV$ has been discussed in the past in \cite{Armillis:2009sm} and it was shown that there is no coupling of this state to 
the vertex for off-shell photons and a virtual massive fermion in the loop. The analysis was performed both for the massive and the massless cases using specific parameterizations of the correlator. We are going to elaborate on this point regarding the way the tensorial limits are performed, not discussed in \cite{Armillis:2009sm}, with the goal of extending these analysis, in the next sections, to the case of the $ATT$. \\
We recall that in order to discuss the presence of a massless particle pole interaction 
in the anomaly vertex we need to perform the limit \eqref{gpp}
and show that it is nonvanishing independently of the virtualities of the other invariants. 
We use as notations  $(s_1,s_2)\equiv (p_1^2, p_2^2)$ and $s=q^2=p_3^3$, assuming various configurations for the external invariants and the value of the fermion mass. We are going to consider two possible ways to approach the $q^2\to 0$ limit, by sending the four-momentum $q^\lambda$ on the light-cone or by performing the soft limit $q^\lambda\to 0$ in all of its components.\\
In order to perform the light-cone limit on $q$ we parameterize the four-momenta as follows 
\beq
p_3\equiv q= q^+ n^+ + q^- n^- \qquad  p_1= p_1^+n^+ + p_1^- n^- + p_\perp
\qquad p_2= p_2^+n^+ + p_2^- n^- - p_\perp
\eeq
with light-cone versors defined by ${(n^\pm)}^2=0 , \, n^+\cdot n^-=1$. In the light cone limit 
$q^+$ is large as well as $p_1^+$ and $p_2^+$, 
while the components 
\beq
q^-=\frac{q^2}{2 q^+} \qquad p_1^-=\frac{p_1^2 + p_\perp^2}{2 p_1^+}\qquad 
p_2^-=\frac{p_2^2 + p_\perp^2}{2 p_2^+}
\eeq
are small. We introduce a scale $\lambda^2\sim p_1^2\sim p_2^2$, with $\lambda \ll \bar{Q}=q^+$.   
In this limit $q_0$ and $q_3$ are both large but fixed, with  $q^-=q_0- q_3$ arbitrarily small. $\bar{Q}$ is assumed of 
any large but {\em fixed} size in the light-cone direction of $n^+$, and the limit $q^2\to 0 $ is performed, for a fixed value of $\bar{Q}$, by taking  
$q^-$, the component in the $n^-$ direction arbitrarily small. For this reason any power of $q^+$ can be dropped if multiplied by $q^-$.\\
The process for extracting the value of the residue of a tensor structure with open indices involves projecting that specific structure onto the light-cone basis. If, during this procedure, a particular component does not vanish, we can conclude that a massless $1/q^2$ pole develops a residue on the light-cone.  \\
It is straightforward to demonstrate that all tensorial structures vanish except for those that include a $1/p_3^2$ pole. This outcome is evident from any parameterization of the correlator. The only case where the residue is 
nonvanishing is for massless vector lines and a massless virtual fermion. We have left to Appendix B.1 more details concerning the technical apsects of this analysis. 
\section{The gravitational chiral anomaly vertex: review }

In this section, we begin our analysis of the \( ATT \) vertex. For simplicity, we are going to set the coupling constant to one. As for the $AVV$, our study starts in the conformal limit, revisiting earlier work that showed how the vertex can be reconstructed from the corresponding anomaly pole in \( CFT_p \) \cite{Coriano:2023gxa}.  
For the \( ATT \) vertex, the conformal realization that we examine in momentum space also encompasses the case of Chern-Simons currents \cite{Dolgov:1988qx}, which generate spin-1 anomalies at first discussed long ago and revisited in more recent studies \cite{delRio:2020cmv,Galaverni:2020xrq}. The form factor decomposition remains valid in both conformal and non-conformal cases, with the mass dependence of the correlator determined via standard perturbation theory. In \cite{Coriano:2023gxa}, it was shown that the conformal solution derived using \( CFT_p \) methods is in full agreement with the perturbative result.  Much like in the \( AVV \) case, a complete determination of the vertex in the conformal limit is also possible for the \( ATT \). Here too, the entire vertex is fixed by the residue at the pole of the longitudinal component of the axial-vector current.  
\\
We start by defining
\begin{align}
\label{ddc}
	T^{\mu_i\nu_i}(p_i)&= t^{\mu_i\nu_i}(p_i)+t_{loc}^{\mu_i\nu_i}(p_i),\\
	\end{align}
having introduced the transverse-traceless ($\Pi$), transverse $(\pi)$ and longitudinal ($\Sigma$) projectors, given respectively by 
\begin{align}
&\Pi^{\mu \nu}_{\alpha \beta}  = \frac{1}{2} \left( \pi^{\mu}_{\alpha} \pi^{\nu}_{\beta} + \pi^{\mu}_{\beta} \pi^{\nu}_{\alpha} \right) - \frac{1}{d - 1} \pi^{\mu \nu}\pi_{\alpha \beta}\label{TTproj}, \\&
	\Sigma^{\mu_i\nu_i}_{\alpha_i\beta_i}=\frac{p_{i\,\beta_i}}{p_i^2}\Big[2\delta^{(\nu_i}_{\alpha_i}p_i^{\mu_i)}-\frac{p_{i\alpha_i}}{(d-1)}\left(\delta^{\mu_i\nu_i}+(d-2)\frac{p_i^{\mu_i}p_i^{\nu_i}}{p_i^2}\right)\Big]+\frac{\pi^{\mu_i\nu_i}(p_i)}{(d-1)}\delta_{\alpha_i\beta_i}\label{Lproj}.
\end{align}
Such decomposition allows to split our correlation function into the following terms
\begin{equation} 
	\label{eq:splitlongttpart}
	\begin{aligned}
		\left\langle T^{\mu_{1} \n_{1}} T^{\mu_{2} \n_2} J_A^{\mu_{3}}\right\rangle=&\left\langle t^{\mu_{1} \n_{1}} t^{\mu_{2}\n_2} j_A^{\mu_{3}}\right\rangle+\left\langle T^{\mu_{1} \n_{1}} T^{\mu_{2}\n_2} j_{5\, l o c}^{\mu_{3}}\right\rangle+\left\langle T^{\mu_{1} \n_{1}} t_{l o c}^{\mu_{2}\n_2} J_A^{\mu_{3}}\right\rangle+\left\langle t_{l o c}^{\mu_{1} \n_{1}} T^{\mu_{2}\n_2} J_A^{\mu_{3}}\right\rangle \\
		&-\left\langle T^{\mu_{1} \n_{1}} t_{l o c}^{\mu_{2}\n_2} j_{A\, l o c}^{\mu_{3}}\right\rangle-\left\langle t_{l o c}^{\mu_{1} \n_{1}} t_{l o c}^{\mu_{2}\n_2} J_A^{\mu_{3}}\right\rangle-\left\langle t_{l o c}^{\mu_{1} \n_{1}} T^{\mu_{2}\n_2} j_{A\, l o c}^{\mu_{3}}\right\rangle+\left\langle t_{l o c}^{\mu_{1} \n_{1}} t_{l o c}^{\mu_{2}\n_2} j_{5\, l o c}^{\mu_{3}}\right\rangle 
	\end{aligned}
\end{equation}
which are constrained by ordinary trace Ward identities of the form 
\begin{equation} \label{ppp}
	\begin{aligned}
		&\delta_{\mu_i\nu_i }\braket{T^{\mu_1\nu_1}(p_1)T^{\mu_2\nu_2}(p_2)J_A^{\mu_3}(p_3)}=0,\qquad \quad&& i=\{1,2\}\\
		&p_{i\mu_i}\,\braket{T^{\mu_1\nu_1}(p_1)T^{\mu_2\nu_2}(p_2)J_A^{\mu_3}(p_3)}=0,\qquad \quad && i=\{1,2\}.
	\end{aligned}
\end{equation}
Notice that the trace identity (the first of the two) will be modified in the presence of mass terms, due to the breaking of conformal symmetry. The second remain invariant, being derived from the conservation of the stress energy tensor. As already mentioned the $TTA$ does not involve any renormalization and there is no trace anomaly. However, due to the presence of several diagrams of different topologies that are separately divergent, one needs a 
regularization scheme. Using the WIs above, we notice that most of the terms in Eq.~\eqref{eq:splitlongttpart} vanish, yielding
\begin{equation} 
	\begin{aligned}
		\left\langle T^{\mu_{1} \n_{1}} T^{\mu_{2} \n_2} J_A^{\mu_{3}}\right\rangle=&\left\langle t^{\mu_{1} \n_{1}} t^{\mu_{2}\n_2} j_A^{\mu_{3}}\right\rangle+\left\langle t^{\mu_{1} \n_{1}} t^{\mu_{2}\n_2} j_{A\, l o c}^{\mu_{3}}\right\rangle .
	\end{aligned}
\end{equation}
The longitudinal anomalous Ward identity is given by

\begin{equation}\label{eq:idwanomlp3}
	p_{3\mu_3}\braket{T^{\mu_1\nu_1}(p_1)T^{\mu_2\nu_2}(p_2)J_A^{\mu_3}(p_3)}= 4 i \, a_2 \, (p_1 \cdot p_2) \left\{ \left[\varepsilon^{\nu_1 \nu_2 p_1 p_2}\left(g^{\mu_1 \mu_2}- \frac{p_1^{\mu_2} p_2^{\mu_1}}{p_1 \cdot p_2}\right) +\left( \mu_1 \leftrightarrow \nu_1 \right) \right] +\left( \mu_2 \leftrightarrow \nu_2 \right) \right\},
\end{equation}
\normalsize
with the rhs of the equation above given by the Fourier transform of the second functional derivative of  the density 
$R\tilde R$.
As in the case of the $AVV$, Eq. \eqref{eq:idwanomlp3} in momentum space is solved by the insertion of a anomaly pole $1/p_3^2$ in the longitudinal sector of $J_A$
\begin{equation} \label{eq:anompolettj}
	\left\langle t^{\mu_{1} \n_{1}} t^{\mu_{2}\n_2} j_{A\, l o c}^{\mu_{3}}\right\rangle= 4i a_2 \frac{p_3^{\mu_3}}{p_3^2} \, (p_1 \cdot p_2) \left\{ \left[\varepsilon^{\nu_1 \nu_2 p_1 p_2}\left(g^{\mu_1 \mu_2}- \frac{p_1^{\mu_2} p_2^{\mu_1}}{p_1 \cdot p_2}\right) +\left( \mu_1 \leftrightarrow \nu_1 \right) \right] +\left( \mu_2 \leftrightarrow \nu_2 \right) \right\}.
\end{equation}
On the other hand, after accounting for the Bose symmetry and all the Schouten identities, the transverse-traceless component of the correlator can be expressed as \cite{Coriano:2023gxa}
\begin{equation}
	\begin{aligned}
		\langle t^{\mu_{1} \nu_{1}}\left({p}_{1}\right)& t^{\mu_{2} \nu_{2}}\left({p}_{2}\right) j_A^{\mu_{3} }\left({p_3}\right)\rangle=\Pi_{\alpha_{1} \beta_{1}}^{\mu_{1} \nu_{1}}\left({p}_{1}\right) \Pi_{\alpha_{2} \beta_{2}}^{\mu_{2} \nu_{2}}\left({p}_{2}\right) \pi_{\alpha_{3}}^{\mu_{3}}\left({p_3}\right) \bigg[
		\\
		&A_1\varepsilon^{p_1\alpha_1\alpha_2\alpha_3}p_2^{\beta_1}p_3^{\beta_2}
		-A_1(p_1\leftrightarrow p_2) \varepsilon^{p_2\alpha_1\alpha_2\alpha_3}p_2^{\beta_1}p_3^{\beta_2}\\
		&+A_2\varepsilon^{p_1\alpha_1\alpha_2\alpha_3}\delta^{\beta_1\beta_2}-
		A_2(p_1\leftrightarrow p_2)\varepsilon^{p_2\alpha_1\alpha_2\alpha_3}\delta^{\beta_1\beta_2}\\
		&+A_3\varepsilon^{p_1p_2\alpha_1\alpha_2}p_2^{\beta_1}p_3^{\beta_2}p_1^{\alpha_3}
		+A_4\varepsilon^{p_1p_2\alpha_1\alpha_2}\delta^{\beta_1\beta_2}p_1^{\alpha_3}
		\bigg],
	\end{aligned}
\end{equation}
where $A_3$ and $A_4$ are antisymmetric under the exchange $(p_1\leftrightarrow p_2)$.
 The solution is \cite{Coriano:2023gxa}

\begin{allowdisplaybreaks}
\begin{equation} \label{eq:pertresultff}
	\begin{aligned}
		A_1&=\frac{ p_2^2}{24 \pi^2 \lambda^4}
		\Bigg\{A_{11} 
		+A_{12} \log\left(\frac{p_1^2}{p_2^2}\right)+A_{13} \log\left(\frac{p_1^2}{p_3^2}\right)
		+A_{14} \, \,C_0(p_1^2,p_2^2,p_3^2)
		\Bigg\}, \\[8pt]
		A_2&=\frac{p_2^2}{48\pi^2\lambda^3}\Bigg\{
		A_{21}
		+A_{22} \log\left(\frac{p_1^2}{p_2^2}\right)+A_{23} \log\left(\frac{p_1^2}{p_3^2}\right)
		+A_{24} \, \, C_0(p_1^2,p_2^2,p_3^2)
		\Bigg\}, \\[8pt]
		A_3&=0,\\[8pt]
		A_4&=0\\[8pt],
	\end{aligned}
\end{equation}
\end{allowdisplaybreaks}
with the K\"{a}llen $\lambda$-function given by
\begin{equation}
	\lambda \equiv \lambda(p_1,p_2,p_3)=\left(p_1-p_2-p_3\right)\left(p_1+p_2-p_3\right)\left(p_1-p_2+p_3\right).\left(p_1+p_2+p_3\right),
	\label{Kallen}
\end{equation}
expressed in terms of the magnitudes of the three momenta. 
In this case, the explicit expression aligns well with perturbation theory, which offers a free field theory realization of the anomalous conformal field theory (CFT). The coefficients \( A_{ij} \) in \eqref{eq:pertresultff} are polynomial functions of the external invariants \( q^2, p_1^2, p_2^2 \) and can be found in \cite{Coriano:2023gxa}.\\
The reconstruction of the interaction highlights the central role of the pole contribution in determining the entire vertex. Importantly, the Ward identities are sufficient to eliminate the apparent divergences in the diagram. Notably, topological contributions do not require any counterterms, provided the remaining conservation Ward identities are imposed on the vertex.
A key point here is that the anomaly pole does not manifest as a residue of the full vertex, as for the $AVV$. We will explore this phenomenon in detail.
\section{Perturbative computation of the massive $\langle TTJ_A \rangle$ correlator}
\begin{figure}[t]
    \centering
    \includegraphics[scale=0.5]{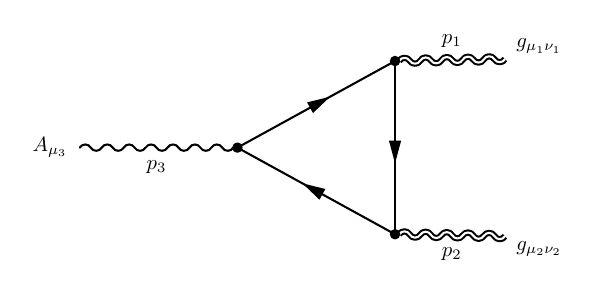}
    \includegraphics[scale=0.5]{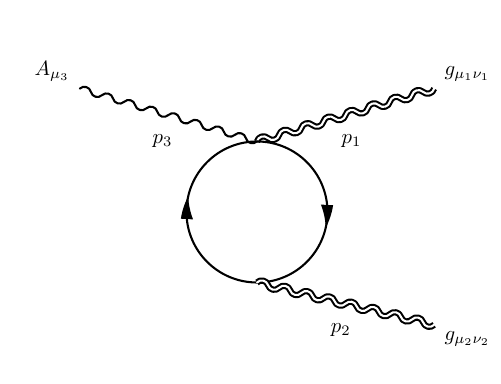}
    \includegraphics[scale=0.5]{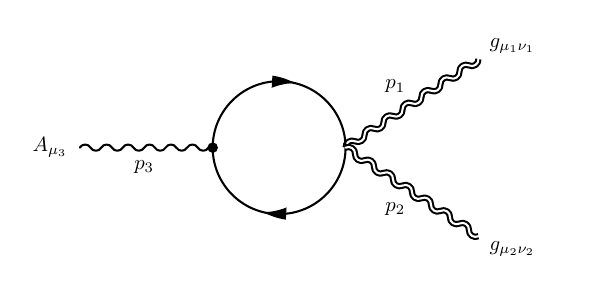}
    \includegraphics[scale=0.5]{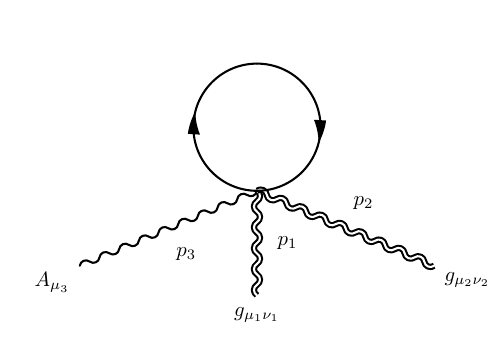}
    \caption{Feynman diagrams representing triangle, bubble, and tadpole topologies in the perturbative computation of the massive correlator.} 
    \label{fig:feynmdiagr}
\end{figure}

The investigation of the sum rule for the anomaly form factors of this vertex requires that we move away from the conformal point, by giving a mass to the fermion. For this reason we turn to the perturbative realization, expanding the analysis presented in \cite{Coriano:2023gxa}. 
Our calculations are carried out in the Breitenlohner-Maison scheme. To streamline the analysis, we transition to Minkowski space, where the generating functional is defined as
\begin{equation}
    e^{i \mathcal{S}[g]} = \int [\mathcal{D} \Phi]\, e^{i S_0[\Phi, g]},
\end{equation}
with the classical action $S_0$ involving a fermionic field in a gravitational and axial gauge field background
\begin{equation}
    S_0 = \int d^d x \,e  \left[\frac{i }{2}e_a^\mu\left( \bar{\psi} \gamma^a (D_\mu \psi) -  (D_\mu \bar{\psi}) \gamma^a \psi\right) -m \bar{\psi}\psi\right],
\end{equation}
where $e_a^\mu$ is the vielbein, $e$ its determinant, and $D_\mu$ the covariant derivative defined as
\begin{equation}
    \begin{aligned}
        D_\mu \psi &= \left(\partial_\mu + i g \gamma_5 B_\mu + \frac{1}{2} \omega_{\mu ab} \Sigma^{ab}\right) \psi, \\
        D_\mu \bar{\psi} &= \left(\partial_\mu - i g \gamma_5 B_\mu - \frac{1}{2} \omega_{\mu ab} \Sigma^{ab}\right) \bar{\psi}.
    \end{aligned}
\end{equation}
Here, $\Sigma^{ab}$ are the generators of the Lorentz group for spin-$\frac{1}{2}$ fields, and $\omega_{\mu ab}$ is the spin connection
\begin{equation}
    \omega_{\mu ab} = e_a^\nu \left(\partial_\mu e_{\nu b} - \Gamma_{\mu\nu}^\lambda e_{\lambda b}\right).
\end{equation}
The Greek indices refer to the curved background, while Latin indices correspond to the local flat basis.
Substituting the explicit form of $\Sigma^{ab}$, the action becomes
\begin{equation}
    S_0 = \int d^d x \, e \left[\frac{i}{2} \bar{\psi} e_a^\mu \gamma^a (\partial_\mu \psi) 
    - \frac{i}{2} (\partial_\mu \bar{\psi}) e_a^\mu \gamma^a \psi 
    - g B_\mu \bar{\psi} e_a^\mu \gamma^a \gamma_5 \psi 
    + \frac{i}{4} \omega_{\mu ab} e_c^\mu \bar{\psi} \gamma^{abc} \psi -m \bar{\psi}\psi\right],
\end{equation}
where $\gamma^{abc} = \{\Sigma^{ab}, \gamma^c\}$.
The corresponding vertices can be computed by functional differentiation of the action with respect to the metric and the gauge field, followed by a Fourier transformation to momentum space. \\
The perturbative computation of the $\langle TTJ_A \rangle$ involves three topologically distinct Feynman diagrams, illustrated in Figure \ref{fig:feynmdiagr}.

\subsection{The form factors for nonzero mass}\label{secffnmastta}
We present the perturbative results of the $TTA$ correlator in the massive case.
The $TTA$ can be decomposed into different sectors. In the massive case, the conformal symmetry is broken and there are additional terms that need to be considered in the decomposition. 
Indeed, the energy-momentum tensor is clearly still conserved, but the correlator exhibits trace terms due to the inclusion of a fermion mass
\begin{equation}
	p_i\left\langle T^{\mu_{1} \n_{1}} T^{\mu_{2} \n_2} J_A^{\mu_{3}}\right\rangle=0, \qquad\qquad
	g_{\mu_i\nu_i}\left\langle T^{\mu_{1} \n_{1}} T^{\mu_{2} \n_2} J_A^{\mu_{3}}\right\rangle\neq 0
	\qquad i=\{1,2\}.
\end{equation}
Therefore, one can write
\begin{equation}
\label{ddf}
	\begin{aligned}
		\left\langle T^{\mu_{1} \n_{1}} T^{\mu_{2} \n_2} J_A^{\mu_{3}}\right\rangle=\left\langle t^{\mu_{1} \n_{1}} t^{\mu_{2}\n_2} j_A^{\mu_{3}}\right\rangle+\left\langle t^{\mu_{1} \n_{1}} t^{\mu_{2}\n_2} j_{A\,\, l o c}^{\mu_{3}}\right\rangle+\left\langle t_{loc}^{\mu_{1} \n_{1}} t^{\mu_{2}\n_2} j_A^{\mu_{3}}\right\rangle+\left\langle t^{\mu_{1} \n_{1}} t^{\mu_{2}\n_2}_{loc} j_A^{\mu_{3}}\right\rangle . 
	\end{aligned}
\end{equation}
We comment on this decomposition for clarity, noting the difference between this expansion, which includes only four terms, and the general one given in \eqref{eq:splitlongttpart}. The suffix ``$loc$'' in the current indicates the longitudinal sector, defined along the direction of momentum \( p_3 \). In the case of the stress-energy tensor, \( t^{\mu\nu}_{loc} \) denotes the sector orthogonal to the transverse-traceless one, according to the decompositions in \eqref{ddc}.  
\\
The conservation of the stress-energy tensor \( T^{\mu\nu} \) in this decomposition leaves, in the massive fermion case, only a trace component in \( t^{\mu\nu}_{loc} \). This trace cannot be accompanied, in the expansion of the correlator, by an operator such as \( j^\mu_{loc} \). The absence of contributions of the form \( \langle t_{loc} \,t\,  j_{loc}\rangle \) can also be understood from symmetry considerations: one cannot construct a tensor of odd parity in the longitudinal sector of the axial-vector current, given the symmetry of \( t \) and the presence of a trace in \( t_{loc} \), with  $j^\mu_{loc} \sim \frac{p^\mu_3}{p_3^2}$.
\\
Similar arguments show that terms of the form \( \langle t_{loc} t_{loc} j_{loc} \rangle \) cannot contribute as parity-odd components of the correlator. Indeed, a perturbative expansion confirms this result
\begin{equation}
	\begin{aligned}
		&g_{\mu_1\nu_1}\,g_{\mu_2\nu_2}\left\langle T^{\mu_{1} \n_{1}} T^{\mu_{2} \n_2} J_A^{\mu_{3}}\right\rangle=0,\\&
		{p_3}_{ \mu_3}\, g_{\mu_i\nu_i}\left\langle T^{\mu_{1} \n_{1}} T^{\mu_{2} \n_2} J_A^{\mu_{3}}\right\rangle =0, \qquad\qquad \quad i=\{1,2\}.
	\end{aligned}
\end{equation}
proving the consistency of the computation. Of all the terms contained in \eqref{ddf}, the anomaly form factor of the $ATT$ is present in the longitudinal component 
\begin{equation}
\label{ffk}
	\left\langle t^{\mu_1 \nu_1} t^{\mu_2 \nu_2} j_{A\, \, l o c}^{\mu_3} \right\rangle = \frac{p_3^{\mu_3} }{p_3^2}\Pi^{\mu_1 \nu_1}_{\alpha_1 \beta_1}(p_1) \Pi^{\mu_2 \nu_2}_{\alpha_2 \beta_2}(p_2) \epsilon^{\alpha_1 \alpha_2 p_1 p_2} \left( \bar F_1 g^{\beta_1 \beta_2} (p_1\cdot p_2) + \bar F_2 p_1^{\beta_2} p_2^{\beta_1} \right).
\end{equation}
This term is determined by the gravitational anomaly, along with the mass corrections. The explicit expressions 
of the form factors \eqref{ffk} are given by 
\small
\begin{equation}
	\begin{aligned}
		\bar F_1&=\frac{1}{24 \pi^2 }+
		\frac{m^2}{2 \pi ^2 \lambda \left(s-s_1-s_2\right) }\bigg\{2 \big[\lambda  m^2+s {s_1} {s_2}\big] {C}_0\left(s,{s_1},{s_2},m^2\right)+Q (s) \big[s ({s_1}+{s_2})-({s_1}-{s_2})^2\big]\\&
		-{s_2} Q ({s_2}) \big[s+{s_1}-{s_2}\big]-{s_1} Q( {s_1}) \big[s-{s_1}+{s_2}\big]+\lambda \bigg\},\\
		\bar F_2&=-\frac{1}{24\pi^2 }+\frac{ m^2}{2 \pi ^2 \lambda ^2 } \bigg\{2 \big[\lambda  \left(m^2 (-s+{s_1}+{s_2})-2 {s_1} {s_2}\right)+3 {s_1} {s_2} \left(({s_1}-{s_2})^2-s ({s_1}+{s_2})\right)\big] {C}_0\left(s,{s_1},{s_2},m^2\right)\\&+{s_1} Q( {s_1}) \big[\lambda+6s_2(s+s_1-s_2)\big]+{s_2} Q ({s_2}) \big[\lambda +6 {s_1} (s-{s_1}+{s_2})\big]-Q (s) \big[12 s {s_1} {s_2}+\lambda  ({s_1}+{s_2})\big]\\&+\lambda  (-s+{s_1}+{s_2})\bigg\}.
	\end{aligned}
\end{equation}
\normalsize
The trace term of the correlator, on the other hand, can be written as
\begin{equation}
	\begin{aligned}
			\left\langle t_{loc}^{\mu_{1} \n_{1}} t^{\mu_{2}\n_2} j_A^{\mu_{3}}\right\rangle\equiv&\, \Sigma^{\mu_1 \nu_1}_{\alpha_1 \beta_1}(p_1) \Pi^{\mu_2 \nu_2}_{\alpha_2 \beta_2}(p_2)\pi^{\mu_3 }_{\alpha_3}(p_3) \left\langle T^{\mu_{1} \n_{1}} T^{\mu_{2} \n_2} J_A^{\mu_{3}}\right\rangle\\& \qquad= \frac{\pi^{\mu_1\nu_1}(p_1)}{3}\Pi^{\mu_2 \nu_2}_{\alpha_2 \beta_2}(p_2)  \pi^{\mu_3 }_{\alpha_3}(p_3)\left\langle T^{\mu}_\mu\, T^{\mu_{2} \n_2} J_A^{\mu_{3}}\right\rangle
	\end{aligned}
\end{equation}
where, in the final step, we used the explicit expression for $\Sigma^{\mu_1 \nu_1}_{\alpha_1 \beta_1}$ given in \eqref{Lproj} along with the conservation of the energy-momentum tensor. This component is purely a trace term and it is transverse with respect to all the momenta. It can be expressed as
\begin{equation}
	\left\langle t_{loc}^{\mu_{1} \n_{1}} t^{\mu_{2}\n_2} j_A^{\mu_{3}}\right\rangle=\frac{\pi^{\mu_1\nu_1}(p_1)}{3}\Pi^{\mu_2 \nu_2}_{\alpha_2 \beta_2}(p_2)  \pi^{\mu_3 }_{\alpha_3}(p_3)\left[\bar F_3\,  \epsilon^{\alpha_2\alpha_3 p_1p_2}p_1^{\beta_{2}}\right]
\end{equation}
where
\small
\begin{equation}
	\begin{aligned}
		&\bar F_3=-\frac{  m^2  s_2 }{2\pi^2 \lambda^2 s} \bigg\{  2 \lambda  s+ 2Q(s) \Big[ 2 \lambda  s + 3  s  \Big( s_1^2 + 6  s_1  s_2 + s_2^2 \Big) + 3 (s_1+s_2)\Big( \lambda - (s_1 - s_2)^2 \Big) \Big] \\&
		- 2 Q(s_1) \Big[  s ( \lambda + 3 s_1 (s_1 + 3  s_2) ) + 3  s_1 ( \lambda - (s_1 - s_2)^2 ) \Big] 
		- 2 Q(s_2) \Big[ s ( \lambda + 3  s_2 (3  s_1 + s_2) ) + 3  s_2 ( \lambda - (s_1 - s_2)^2 ) \Big] \\
		&\qquad +\Big[
		\lambda s \left(   4m^2 + 3 (s_1 + s_2) \right) + 24  s s_1  s_2 (s_1 + s_2) 
		- \left( (s_1 - s_2)^2 - \lambda \right) \left( \lambda + 12  s_1  s_2 \right)\Big] C_0(s, s_1, s_2, m^2) 
		\bigg\}.
	\end{aligned}
\end{equation}
\normalsize
Finally, the transverse-traceless part can be expressed as
\begin{equation}
	\begin{aligned}
		\langle t^{\mu_{1} \nu_{1}}\left({p}_{1}\right)& t^{\mu_{2} \nu_{2}}\left({p}_{2}\right) j_A^{\mu_{3} }\left({p_3}\right)\rangle=\Pi_{\alpha_{1} \beta_{1}}^{\mu_{1} \nu_{1}}\left({p}_{1}\right) \Pi_{\alpha_{2} \beta_{2}}^{\mu_{2} \nu_{2}}\left({p}_{2}\right) \pi_{\alpha_{3}}^{\mu_{3}}\left({p_3}\right) \bigg[
		\\
		&A_1\varepsilon^{p_1\alpha_1\alpha_2\alpha_3}p_2^{\beta_1}p_3^{\beta_2}
		-A_1(p_1\leftrightarrow p_2) \varepsilon^{p_2\alpha_1\alpha_2\alpha_3}p_2^{\beta_1}p_3^{\beta_2}\\
		&+A_2\varepsilon^{p_1\alpha_1\alpha_2\alpha_3}\delta^{\beta_1\beta_2}-
		A_2(p_1\leftrightarrow p_2)\varepsilon^{p_2\alpha_1\alpha_2\alpha_3}\delta^{\beta_1\beta_2}\\
		&+A_3\varepsilon^{p_1p_2\alpha_1\alpha_2}p_2^{\beta_1}p_3^{\beta_2}p_1^{\alpha_3}
		+A_4\varepsilon^{p_1p_2\alpha_1\alpha_2}\delta^{\beta_1\beta_2}p_1^{\alpha_3}
		\bigg]
	\end{aligned}
\end{equation}
where
\begin{equation}
	\begin{aligned}
		A_1 &=- \frac{ s_2}{24 \pi^2 \lambda^4}\Big[\bar{A}_{11} Q(s_1)+\bar{A}_{12}Q(s_2)+\bar{A}_{13}Q(s)+\bar{A}_{14}C_0(s, s_1, s_2, m^2)+\bar{A}_{15}\Big],\\
		A_2 &=-\frac{  s_2}{48 \pi^2 \lambda^3} \Big[\bar{A}_{21} Q(s_1)+\bar{A}_{22}Q(s_2)+\bar{A}_{23}Q(s)+\bar{A}_{24}C_0(s, s_1, s_2, m^2)+\bar{A}_{25}\Big],\\
		A_3 &=0,\\
		A_4 &=0.
	\end{aligned}
\end{equation}
The complete expressions of these contributions can be found in Appendix \ref{transverse}.

\section{The sum rule of the $ATT$ for on-shell and off-shell gravitons }
Before discussing the sum rule of the gravitational anomaly form factor in full generality, we consider the case in which the two gravitons are on-shell.  
 The on-shell expression of the $ATT$ takes a form close to the $AVV$.  
 A direct perturbative computation shows that also in this case the only tensor structure present is the longitudinal one, proportional to $R\tilde{R}$ 
\begin{equation}
\label{all}
	\begin{aligned}
		\langle T^{\mu_1 \nu_1} T^{\mu_2 \nu_2} J^{\mu_3}_A\rangle&=\frac{1}{96 \pi^2}\left[1+ \frac{12m^2}{p_3^2} \bigg(1+2 m^2 {C}_0\left(p_3^2,0,0, m^2\right)\bigg)\right]\\&\times \frac{p_3^{\mu_3}}{p_3^2}\bigg\{\bigg[\varepsilon^{\nu_1 \nu_2 p_1 p_2}\left(\left(p_1 \cdot p_2\right)g^{\mu_1 \mu_2}-{p_1^{\mu_2} p_2^{\mu_1}} \right)+\left(\mu_1 \leftrightarrow \nu_1\right)\bigg]+\left(\mu_2 \leftrightarrow \nu_2\right)\bigg\} .
	\end{aligned}
\end{equation}
 To investigate the presence of a pole in the corresponding form factor, the \( C_0(s, 0, 0, m^2) \) integral needs to be expanded around $s=0$. We can use the expression \eqref{exp0} or, equivalently, we can rewrite it in the form 
\begin{equation}
	C_0(s, 0, 0, m^2) = \frac{\log^2 \left( \frac{\sqrt{s \left( s - 4m^2 \right)} + 2m^2 - s}{2m^2} \right)}{2s}.
\end{equation}
To better understand the behavior of \( C_0(s, 0, 0, m^2) \) near \( s = 0 \), we perform a series expansion. Expanding \( C_0(s, 0, 0, m^2) \) around \( s = 0 \), as in \eqref{exp1}, allows for a direct computation that confirms the absence of a pole at \( s = 0 \).  
Substituting this expansion into \eqref{all} and simplifying the expression reveals that the \( {1}/{s} \) and \( {1}/{s^2} \) terms cancel out. Specifically, the remaining contributions are finite and free of singularities, explicitly demonstrating the cancellation mechanism. Consequently, in the on-shell spectral function, only the cut contribution remains, in complete analogy with the \( AVV \) correlator.\\
  \begin{figure}
	\centering
	\begin{tikzpicture}
	
	\draw[thick, ->] (-1,0) -- (7,0) node[right] {$\text{Re}(s)$};
	\draw[thick, ->] (0,-2) -- (0,4) node[above] {$\text{Im}(s)$};
	
	\filldraw[black] (0,0) circle (3pt) node[below left] {$0$};
	
	\draw[thick, decorate, decoration={zigzag, amplitude=2pt, segment length=4pt}] (2,0) -- (7.0,0);
	\node at (2, -0.3) {$4m^2$};
	
	\draw[very thick, ->, red] (0,0) -- (0,1.5);     
	\draw[very thick, ->, black] (0,0) -- (0,-2.5); 
	
	\draw[thick] (3.5,0.2) -- (3.7,-0.2); 
	\draw[thick] (3.7,0.2) -- (3.5,-0.2); 
	\node[below] at (3.6,-0.3) {$s_-$};   
	
	\draw[thick] (4.5,0.2) -- (4.7,-0.2); 
	\draw[thick] (4.7,0.2) -- (4.5,-0.2); 
		\node[below] at (4.6,-0.3) {$s_+$};   
	\draw[very thick, ->, black] (6.2,0.0) -- (6.2,-1.5); 
         \node[below] at (6.0,-0.3) {$\bar s$}; 
        \end{tikzpicture}
	\caption{The spectral density of the off-shell gravitational anomaly form factor in the complex \( s\equiv q^2 \)-plane. Shown are the two spikes at \(s=0\), the anomalous thresholds at $s_{\pm}=(\sqrt{s_1}\pm \sqrt{s_2})^2$ and the pole at $s=\bar{s}=s_1+s_2$. In this case there is a cancelation in the sum rule between the integral of the density over the cut, the black spike at $s=0$, of variable strength, and the black spike at $s=\bar{s}$.   } 
\label{figgX}
\end{figure}
\noindent
We are going to expand on these analysis, showing the presence of a sum rule in the most general case, extending the approach presented for the $AVV$ in the former sections. 
The conformal description of the off-shell $(s_1, s_2 \neq 0, m = 0)$ $\langle TTJ_A\rangle$ or $\langle TT J_{CS}\rangle $ correlators is identical, as the only distinction between the two scenarios lies in the numerical value of the anomaly. In the conformal limit, the pole is inherently present, as demonstrated in \eqref{eq:anompolettj}. In the massive case, the same form factor can be computed perturbatively using the diagrams shown in Fig. \ref{fig:feynmdiagr}. Similarly, in this case, the perturbative computation does not necessitate any counterterms; however, an explicit regularization scheme is required to handle the $\gamma_5$ matrix within the loops. For this purpose, we have employed the Breitenlohner-Maison scheme for $\gamma_5$. We have presented the results of the computation in Section \ref{secffnmastta}. Specifically, the structure of the correlator in the longitudinal part, which defines the anomaly form factor in this context, can be written as
\begin{equation}
\label{rr1}
	\langle t^{\mu_1 \nu_1} t^{\mu_2 \nu_2} j^{\mu_3}_{A\,loc} \rangle 
	= (\bar F_1-\bar F_2) \frac{p^{\mu_3}_3}{4\, p_3^2} 
	\left\{
	\left[
	\varepsilon^{\nu_1 \nu_2 p_1 p_2} 
	\left(
	p_1 \cdot p_2\,g^{\mu_1 \mu_2} - p_1^{\mu_2} p_2^{\mu_1}
	\right)
	+ (\mu_1 \leftrightarrow \nu_1)
	\right]
	+ (\mu_2 \leftrightarrow \nu_2)
	\right\} \nn
\end{equation}

\begin{equation}
	+ (\bar F_1+\bar F_2)\,\, \frac{p^{\mu_3}_3}{4\, p_3^2} 
	\left\{
	\left[
	\varepsilon^{\nu_1 \nu_2 p_1 p_2} 
	\left(
	p_1 \cdot p_2\,g^{\mu_1 \mu_2} + p_1^{\mu_2} p_2^{\mu_1}
	\right)
	+ (\mu_1 \leftrightarrow \nu_1)
	\right]
	+ (\mu_2 \leftrightarrow \nu_2)
	\right\}\nn
\end{equation}
\begin{equation}
	+ (\bar F_1+\bar F_2)\frac{(p_1 \cdot p_2)}{p_1^2p_2^2} \frac{p^{\mu_3}_3}{4\,p_3^2} 
	\left\{
	\left[
	\varepsilon^{\nu_1 \nu_2 p_1 p_2} 
	\left(
	(p_1 \cdot p_2)\,p_1^{\mu_1}p_2^{\mu_2} - p_2^2\,p_1^{\mu_1} p_1^{\mu_2}-p_1^2\, p_2^{\mu_1} p_2^{\mu_2}
	\right)
	+ (\mu_1 \leftrightarrow \nu_1)
	\right]
	+ (\mu_2 \leftrightarrow \nu_2)
	\right\}
\end{equation}
where the first contribution, proportional to $\bar F_1-\bar F_2$  is related to the second functional derivative of the anomaly $R\tilde{R}$ performed with respect to the metric and defines the anomaly form factor 

\begin{align}
\label{ATT}
	\phi_{ATT}=\frac{\bar F_1-\bar F_2}{s}=\,\,&\frac{m^2 s_1 \left(s_2 \left(s_1^2-s^2\right)+s_2^2 (5 s+s_1)+(s_1-s)^3-3 s_2^3\right) B_0\left(s_1,m^2\right)}{\pi ^2 s (s-s_1-s_2) \lambda^2}\nonumber\\&+\frac{ m^2 s_2 \left(s_2 \left(3 s^2+s_1^2\right)+s_2^2 (s_1-3 s)-(s-s_1)^2 (s+3 s_1)+s_2^3\right) B_0\left(s_2,m^2\right)}{\pi ^2 s (s-s_1-s_2) \lambda^2}\nonumber\\&+\frac{ m^2 K_1(s,s_1,s_2) B_0\left(s,m^2\right)}{\pi ^2  s(s-s_1-s_2) \lambda^2}+\frac{ m^2 K_2(s,s_1,s_2) C_0\left(s,s_1,s_2,m^2\right)}{\pi ^2  s(s-s_1-s_2) \lambda^2}\nonumber\\& \frac{ m^2 \left(s^2-2 s (s_1+s_2)+s_1^2+s_2^2\right)}{\pi ^2s(s-s_1-s_2) \lambda}+ \frac{1}{12 \pi ^2 s}.
\end{align}
Notice that this form factor, at large $s$ behaves as 
\beq
\phi_{ATT} \sim \frac{1}{12 \pi ^2 s} + O(\frac{m^2}{s^2}\log\frac{s}{m^2}),
\eeq
similarly to the the behaviour encountered in the case of the $AVV$, showing the dominance of the anomaly at large $q^2$. \\
The second and the third terms in \eqref{rr1}, which are related to $\bar F_1 +\bar F_2$
\begin{align}\label{eqqq}
	\bar F_1+\bar F_2=\,\,&\frac{2 m^2 s_1s_2 \left(s^2+s (s_1-2 s_2)-2 s_1^2+s_1 s_2+s_2^2\right) B_0\left(s_1,m^2\right)}{\pi ^2  (s-s_1-s_2) \lambda^2}\nonumber\\&+\frac{2  m^2  s_1s_2 \left(s_2 (s+s_1)+(s-s_1)^2-2 s_2^2\right) B_0\left(s_2,m^2\right)}{\pi ^2 s (s-s_1-s_2) \lambda^2}\nonumber\\&+\frac{2  m^2   s_1s_2\left(-2 s^2+s (s_1+s_2)+(s_1-s_2)^2\right) B_0\left(s,m^2\right)}{\pi ^2 (s-s_1-s_2) \lambda^2}\nonumber\\&+\frac{ m^2  s_1s_2
	K_3(s,s_1,s_2)\, C_0\left(s,s_1,s_2,m^2\right)}{\pi ^2  (s-s_1-s_2) \lambda^2}\nonumber\\&-\frac{2 \, m^2  s_1s_2}{\pi ^2  (s-s_1-s_2) \lambda}
\end{align}
are generated by the breaking of conformal symmetry, 
vanishing in the massless case and/or for on-shell gravitons $(s_1=s_2=0)$ and after contraction with the graviton polarizers. 
In equation \eqref{eqqq} the expression of $K_1$, $K_2$ and $K_3$ is given by
\begin{align}
	K_1(s,s_1,s_2)=\,\,&s_2 (s-s_1) \left(s^2+3 s s_1-2 s_1^2\right)+s_2^3 (3 s+2 s_1)\nonumber\\&-s_2^2 (s+s_1) (3 s+2 s_1)+s_1 (s-s_1)^3-s_2^4
\end{align}

\begin{align}
	K_2(s,s_1,s_2)=\,\,&2 m^2 \lambda \left(s^2-2 s (s_1+s_2)+s_1^2+s_2^2\right)\nonumber\\&+s_1 s_2 \left(3 s^3-5 s^2 (s_1+s_2)+s (s_1+s_2)^2+(s_1-s_2)^2 (s_1+s_2)\right)
\end{align}
\begin{align}
	K_3(s,s_1,s_2)=\,\,&s_2 \left(8 m^2 (s+s_1)+s^2-6 s s_1+s_1^2\right)+s_2^2 \left(-4 m^2+s+s_1\right)\nonumber\\&-(s-s_1)^2 \left(4 m^2+s+s_1\right)-s_2^3. 
\end{align}
We are going to show that the combination $\bar F_1-\bar F_2$ defines the correct expression of the anomaly form factor and we will verify the presence of a sum rule also in this case. Notice that the contribution proportional to $\bar F_1 +\bar F_2$, as we have mentioned, is proportional to the virtualities of the two gravitons and vanishes as we require these to be on-shell (see Eq$.$ \eqref{all}). 

\subsection{Derivation of the general sum rule for the $ATT$}
The anomaly form factor is associated with the first structure proportional to $\bar F_1 -\bar F_2$ in \eqref{rr1}. The corresponding tensor structure is the off-shell expression of the second functional derivative of $R\tilde{R}$, Fourier transformed to momentum space. The second contribution, proportional to $\bar F_1 +\bar F_2$, as we have just discussed, vanishes for on-shell-gravitons and is generated 
by the explicit breaking of the chiral symmetry for a nonzero mass parameter.  \\
Coming to the spectral function of the form factor in \eqref{ATT}, we separate the discontinuity into the pole contributions plus the continuum, in the form 
\begin{equation}
	\text{disc}(\phi_{ATT})=\text{disc}(\phi_{ATT})_{cont}+\text{disc}(\phi_{ATT})_{pole},
\end{equation}
with the discontinuity over the continuum given by 
\begin{align}
\label{cont}
	\text{disc}(\phi_{ATT})_{cont}=\frac{ m^2 K_1(s,s_1,s_2)\, \text{disc}\, B_0\left(s,m^2\right)}{\pi ^2  s(s-s_1-s_2) \lambda^2}+\frac{g m^2 K_2(s,s_1,s_2) \,\text{disc}\,C_0\left(s,s_1,s_2,m^2\right)}{\pi ^2  s(s-s_1-s_2) \lambda^2}.
\end{align}
The discontinuity of $C_0$ is given by \eqref{d2}, while 
\begin{equation}
	\textrm{disc}\,{B_0}(s,m^2)=2\p i \sqrt{1-\frac{4m^2}{s} }\,\,\,
	\theta(s-4m^2).
\end{equation}
The pole contributions split at separate locations
\begin{align}
	\text{disc}(\phi_{ATT})_{pole}=\text{disc}(\phi_{ATT})_{0}+\text{disc}(\phi_{ATT})_{s_1+s_2}+\text{disc}(\phi_{ATT})_{s_-}+\text{disc}(\phi_{ATT})_{s_+},
\end{align}
where we have indicated the points corresponding to the zeroes of the $\lambda$ function as
\beq
s_\pm=(\sqrt{s_1} \pm \sqrt{s_2})^2.
\eeq
The pole at $s=0$ is split into two spikes 
\begin{align}
\label{Ym}
	\text{disc}(\phi_{ATT})_{0}=2\p i&\frac{\delta(s)}{12 \pi ^2 }-\frac{2\p i\delta(s)}{12 \pi ^2 (s_1-s_2)^3 (s_1+s_2)}\Y(s,s_1,s_2,m^2)
\end{align}
where 

\begin{align}
\Y(s,s_1,s_2,m^2)\equiv&\biggl(12 m^2 \bigl(-\left((s_1-s_2) \left(s_1^2+s_2^2\right){B}_0\left(s,m^2\right)\right)\nonumber\\
&+s_1 \left(s_1^2+2 s_1 s_2+3 s_2^2\right) {B}_0\left(s_1,m^2\right)-s_2 \left(3 s_1^2+2 s_1 s_2+s_2^2\right){B}_0\left(s_2,m^2\right)\nonumber\\
&+(s_1-s_2) \left(2 m^2 \left(s_1^2+s_2^2\right)+s_1 s_2 (s_1+s_2)\right) C_0\left(s,s_1,s_2,m^2\right)\bigl)\nonumber\\ 
& +12 m^2 \left(s_1^2+s_2^2\right) (s_1-s_2)\biggl).
\end{align}
We have outlined the various contributions in \figref{figgX}. The "black spike"  at \( s = 0 \), which has a constant strength, corresponds to the first term on the right-hand side of \eqref{Ym}. The second contribution, which depends on the external invariants, is represented by the "red spike" in the same figure.  \\
For \( s > 4m^2 \), across the cut, we encounter two apparent poles that, as we will demonstrate, do not contribute to the discontinuity. Additionally, there is a pole at \( s = s_1 + s_2 \), which can lie either above or below the cut (i.e., for \( s > 4m^2 \) or \( s < 4m^2 \)), depending on the values of \( s_1 \) and \( s_2 \).  
Finally, we observe that in the limit \( s \to 0 \)
\begin{align}
\lim_{s\to 0} \Y(s,s_1,s_2)= \Y(0,s_1,s_2,m^2)\equiv&\biggl(12 m^2 \bigl(-\left((s_1-s_2) \left(s_1^2+s_2^2\right){B}^R_0\left(0,m^2\right)\right)\nonumber\\
&+s_1 \left(s_1^2+2 s_1 s_2+3 s_2^2\right){B}_0^R\left(s_1,m^2\right)-s_2 \left(3 s_1^2+2 s_1 s_2+s_2^2\right){B}_0^R\left(s_2,m^2\right)\nonumber\\
&+(s_1-s_2) \left(2 m^2 \left(s_1^2+s_2^2\right)+s_1 s_2 (s_1+s_2)\right){C}_0\left(0,s_1,s_2,m^2\right)\bigl)\nonumber\\ 
& +12\, m^2 \left(s_1^2+s_2^2\right) (s_1-s_2)\biggl)  < \infty
\end{align}
 \(\Y\) remains finite across the entire light-cone.

\subsection{Cancellation of the anomalous thresholds and the sum rule}
\label{trash}
 The expression for the spectral density
 \begin{equation}
 	\Phi_{\text{ATT}}=\frac{F_1 - F_2}{s}
 \end{equation} 
 can be separated into distinct components, representing contributions from regular kinematical thresholds as well as potential anomalous thresholds
\begin{equation}
	\Phi_{\text{ATT}} = \Phi_{\text{ATT}}^{\text{reg}} + \Phi_{\text{ATT}}^{(\l)},
\end{equation}
where $\Phi_{\text{ATT}}^{\text{reg}}$ corresponds to regular kinematical thresholds and $\Phi_{\text{ATT}}^{\l}$ accounts for possible anomalous thresholds. The regular part, $\Phi_{\text{ATT}}^{\text{reg}}$, includes contributions from standard kinematical thresholds located at
\begin{equation}
	q^2 = 0, \quad q^2 = s_1 + s_2, \quad \text{and} \quad q^2 = 4m^2,
\end{equation}
which manifest in the form of simple poles and cuts. These thresholds arise from well-understood physical processes. In contrast, the anomalous part, $\Phi_{\text{ATT}}^{\text{anom}}$, emerges under special conditions when the kinematical factor $\lambda = 0$ which could be present in this computation. Given that
\begin{equation}
\lambda = (s - s_-)(s - s_+),
\end{equation}
the dispersive representation of  $ \Phi_{\text{ATT}}$ may allow terms of the form
\begin{align}
\textrm{disc} \,\Phi_{\text{ATT}} = &c_0 \delta(s) + c_1(s,s_1, s_2) \theta(s - 4m^2) H(s) + c_2(s,s_1, s_2) \delta(s - s_1 - s_2)\nn\\&
+ c_3(s, s_1, s_2) \delta'(s - s_-) + c_4 \delta'(s - s_+) + c_5 \delta(s - s_-) + c_6 \delta(s - s_+),
\end{align}
where the derivative of the $\delta$-function corresponds to higher order poles at $\lambda = 0$.
Indeed, the general structure of $\Phi_{\text{ATT}}$ is
\begin{align}
	\Phi_{\text{ATT}} =&  \frac{c_0(s,s_1,s_2,m^2)}{s} + \frac{c_1(s,s_1,s_2,m^2)}{s} + \text{(continuum terms)} + \frac{c_2(s,s_1,s_2,m^2)}{s - \bar{s}} + \frac{c_3(s,s_1,s_2,m^2)}{(s - s_-)^2} \nn\\&+ \frac{c_4(s,s_1,s_2,m^2)}{(s - s_+)^2} + \frac{c_5(s,s_1,s_2,m^2)}{s - s_-} + \frac{c_6(s,s_1,s_2,m^2)}{s - s_+},
\end{align}

where $\bar{s} = s_1 + s_2$ and, $c_3$ and $c_4$ define possible residues at the double poles of $\lambda^2$. These could, in principle, introduce subtractions that can be analyzed as follows. 
For example, at $s = s_-$, we have
\begin{equation}
	\Phi_{\text{ATT}}^{(\l)}(q^2 = s_-) = \frac{1}{2\pi i} \int_0^\infty \frac{c_3(s, s_1, s_2, m^2) \delta'(s - s_-)}{(s - q^2)} \, ds
	+ \frac{1}{2\pi i} \int_0^\infty \frac{c_5(s, s_1, s_2, m^2) \delta(s - s_-)}{(s - q^2)} \, ds,
\end{equation}
that simplifies to
\begin{equation}
	{2\pi i}\,\Phi_{\text{ATT}}^{(\l)}(q^2 = s_-)= -\frac{d}{ds} \left( \frac{c_3(s, s_1, s_2, m^2)}{(s - q^2)} \right) \bigg|_{s = s_-} - \frac{c_5(s_-, s_1, s_2, m^2)}{s_- - q^2}.
\end{equation}
A similar analysis gives 
\begin{equation}
	{2\pi i}\,\Phi_{\text{ATT}}^{(\lambda)}(q^2 = s_+) =  -\frac{d}{ds} \left( \frac{c_4(s, s_1, s_2, m^2)}{(s - q^2)} \right) \bigg|_{s = s_+} - \frac{c_6(s_+, s_1, s_2, m^2)}{s_+ - q^2}.
\end{equation}
For the evaluation of $\phi_{\text{ATT}}^{(\lambda)}(q^2 = s_\pm)$, we need to compute the explicit expression of the scalar integral $C_0$ at these kinematical points. We use the expressions of the scalar integral $C_0$ at these points. They are given by 
\begin{align}
C_0(s_-,s_1,s_2,m^2)=&\frac{1}{{s_1}{s_2}\, s_-^2}\Biggl(\sqrt{{s_1}} \Biggl(\sqrt{{s_2} \left({s_2}-4 m^2\right)} \left(\sqrt{{s_1}}-\sqrt{{s_2}}\right)\times \nn\\ &
\times \log \left(\frac{\sqrt{{s_2} \left({s_2}-4 m^2\right)}+2 m^2-{s_2}}{2 m^2}\right)\nonumber\\&+\sqrt{{s_2}} \sqrt{\left(-2 \sqrt{{s_1}} \sqrt{{s_2}}+{s_1}+{s_2}\right) \left(-4 m^2-2 \sqrt{{s_1}} \sqrt{{s_2}}+{s_1}+{s_2}\right)} \log \left(S_1\right)\Biggl)\nonumber\\&+\sqrt{{s_2}} \sqrt{{s_1} \left({s_1}-4 m^2\right)} \left(\sqrt{{s_2}}-\sqrt{{s_1}}\right) \log \left(\frac{\sqrt{{s_1} \left({s_1}-4 m^2\right)}+2 m^2-{s_1}}{2 m^2}\right)\Biggl)
\end{align}
and 
\begin{align}
	C_0( s_+,s_1,s_2,m^2)=&\frac{1}{{s_1}{s_2} s_+^2}\Biggl(\sqrt{{s_1}} \Biggl(\sqrt{{s_2} \left({s_2}-4 m^2\right)} \left(\sqrt{{s_1}}+\sqrt{{s_2}}\right)\times \nn \\&
\times \log \left(\frac{\sqrt{{s_2} \left({s_2}-4 m^2\right)}+2 m^2-{s_2}}{2 m^2}\right)\nonumber\\&-\sqrt{{s_2}} \sqrt{\left(\sqrt{{s_1}}+\sqrt{{s_2}}\right)^4-4 m^2 \left(\sqrt{{s_1}}+\sqrt{{s_2}}\right)^2} \log \left(S_2\right)\Biggl)\nonumber\\&+\sqrt{{s_2}} \sqrt{{s_1} \left({s_1}-4 m^2\right)} \left(\sqrt{{s_1}}+\sqrt{{s_2}}\right) \log \left(\frac{\sqrt{{s_1} \left({s_1}-4 m^2\right)}+2 m^2-{s_1}}{2 m^2}\right)\Biggl), 
\end{align}

where
\begin{equation}
	S_1=\frac{\sqrt{\left(-2 \sqrt{{s_1}} \sqrt{{s_2}}+{s_1}+{s_2}\right) \left(-4 m^2-2 \sqrt{{s_1}} \sqrt{{s_2}}+{s_1}+{s_2}\right)}+2 m^2-\left(\sqrt{{s_1}}-\sqrt{{s_2}}\right)^2}{2 m^2}
\end{equation}
and 
\begin{equation}
	S_2=\frac{\sqrt{\left(\sqrt{{s_1}}+\sqrt{{s_2}}\right)^4-4 m^2 \left(\sqrt{{s_1}}+\sqrt{{s_2}}\right)^2}+2 m^2-\left(\sqrt{{s_1}}+\sqrt{{s_2}}\right)^2}{2 m^2}.
\end{equation}
Using these relations one can show by a long computation that 

\begin{equation}
	\Phi_{\text{ATT}}^{(\l)}(q^2 = s_\pm) \equiv 0,
\end{equation}
and hence 
\begin{equation}
	\Phi_{\text{ATT}}^{(\l)}=0.
\end{equation}
Therefore, the only pole to consider at finite $s$ is the one located at $s= s_1+s_2$.
This contributes to the sum rule with a localized density that is given by 
\begin{align}
\label{bo}
	\text{disc}(\phi_{ATT})_{s_1+s_2}=&\frac{2\p i\, m^2\delta(s-s_1-s_2)}{4 \pi ^2 (s_1+s_2)}  \biggl(-2 {B}_0\left(s_1+s_2,m^2\right)+{B}_0\left(s_1,m^2\right)\nn\\&+{B}_0\left(s_2,m^2\right)+\left(4 m^2-s_1-s_2\right){C}_0\left(s_1,s_2,s_1+s_2,m^2\right)+2\biggl).
\end{align}
Notice that the UV divergences in the $B_0$'s in the expression above cancel automatically, with rule $-2 +1 +1$ in the counting of the $1/\epsilon_{UV}$ poles in each of these self-energy contributions. Therefore we can replace in \eqref{bo} each $B_0$ by $B_0^R$ with no change. No IR singularities are encountered unless we consider special values where any of the external invariants vanishes. 
\\
In order to compute the sum rule for the longitudinal part of the $ATT$ correlator we can use the same technique used before for the $AVV$. Therefore, as in the case of the anomaly form factor of the $AVV$,  we apply the inverse Laplace transform on both sides of the dispersive representation of the form factor (Cauchy's relation) to obtain
	\begin{equation}
	\mathcal{L}^{-1}\Bigl\{   \int_0^{\infty} \frac{\Delta \phi_{ATT}}{s-\bar{s}}  \,ds \Bigl\} (t) = \frac{1}{\pi} \int_0^{\infty} 
	\Delta \phi_{ATT} \,\,e^{{{s}}t}  \,ds,	
\end{equation}
where 
\begin{equation}
	\Delta \phi_{ATT}\equiv \frac{1}{2 i}\text{disc}\, \phi_{ATT}
\end{equation}
and the integral of the spectral density will be obtained by performing the $t\to 0$ limit
\beq	\label{sumr}
\lim_{t\rightarrow 0}\mathcal{L}^{-1}\{\phi_{ATT}\}(t)  =\frac{1}{\pi}\int_0^{\infty} \Delta \phi_{ATT}  \,ds.  
\eeq
The techniques used to compute the inverse Laplace transform is the same described in the previous case. We have used the parametric form of the 2-points and the 3-points scalar function, exchanging the integration with the inverse Laplace transform, and once the limit on $t$ has been performed, we obtain the result of the integral in equation \eqref{sumr}
\begin{equation}
\label{ag}
	\frac{1}{\pi}\int_0^{\infty} \Delta \Phi_{ATT}  \,ds = a_g
\end{equation}
with $a_g$ denoting the gravitational anomaly. The pattern of cancelation, in this case, is between the continuum, the pole at 
$s=\bar s=s_1 + s_2$ and the pole at $s=0$ giving the identity 
\beq
\label{eq}
\int_0^\infty \frac{1}{2 i} \left(\textrm{disc}\Phi_{ATT\, 0}(s,s_1,s_2,m^2) + \textrm{disc}\Phi_{ATT\, s_1+s_2}(s,s_1,s_2,m^2)\right)=\int_{4 m^2}^\infty 
ds \,  \frac{1}{2 i} \textrm{disc}\Phi_{ATT}(s,s_1,s_2,m^2)
\eeq 
where 
\beq
\int_0^\infty \frac{1}{2 i} \left(\textrm{disc}\Phi_{ATT\, 0}(s,s_1,s_2,m^2) + \textrm{disc}\Phi_{ATT\, s_1+s_2}\right) =
\nonumber
\eeq
\beq
\frac{1}{12 \pi} - \frac{1}{12 \pi  (s_1-s_2)^3 (s_1+s_2)}\Y(0,s_1,s_2,m^2)
\nonumber
\eeq
\beq
+ \frac{ m^2}{4 \pi  (s_1+s_2)} \biggl(-2 {B}^R_0\left(s_1+s_2,m^2\right)+{B}^R_0\left(s_1,m^2\right)
\nonumber
\eeq
\beq
+{B}^R_0\left(s_2,m^2\right)+\left(4 m^2-s_1-s_2\right) {C}_0\left(s_1,s_2,s_1+s_2,m^2\right)+2\biggr),
\label{qw1}
\eeq
while the rhs of \eqref{eq} is given by integrating \eqref{cont}. \eqref{ag} is the result of this cancelation. The 
proof of this identity can be obtained by using the parametric representation of the integrals 
and the Laplace/Borel method, followed by complete numerical checks. 

\section{Conclusions}
In this work, we have examined both local and nonlocal formulations of axion interactions, highlighting their distinctions and limitations—particularly in relation to anomaly dynamics. The local interaction, derived under specific symmetry assumptions, is well-suited to contexts such as QCD, where an axion mass emerges at the confinement scale. In such scenarios, the conceptual constraints identified in our analysis can be effectively bypassed.\\
A central challenge lies in identifying the conditions under which local and nonlocal formulations become equivalent. From the perspective of the one-particle-irreducible (1PI) effective action, the anomaly form factors associated with nonlocal interactions are constrained by perturbative sum rules. These sum rules govern the dynamics of virtual states and serve as key signatures of anomaly-induced processes. Importantly, we have demonstrated that these rules remain valid even in the presence of gravitational anomalies and for arbitrary fermion masses—despite being violated by the local action.\\
The sum rules differentiate between contributions from particle poles and branch cuts, offering a comprehensive framework for understanding the off-shell behavior of the interaction. While the local formulation captures on-shell physics, it fails to reflect this richer structure. \\
We have illustrated the nontrivial interplay between poles, branch cuts, offering a different perspective on the emergence of such interactions, relevant in the context of QCD factorization and the proton spin, which has been recently debated in several recent works \cite{Bhattacharya:2022xxw,Tarasov:2020cwl,Castelli:2024eza}. 
In the conformal limit, anomaly interactions impose nontrivial constraints on both longitudinal and transverse sectors via conformal Ward identities. Spectral density flows further reveal a redistribution of contributions between the continuum and massless poles.
This study significantly extends previous investigations by providing explicit off-shell spectral density representations of the chiral anomaly form factors for both correlators.  
For massless fermions, the anomaly and particle poles coincide only on-shell; off-shell, anomaly-induced interactions behave differently. Notably, no particle pole persists for massive fermions or vector bosons. Nonetheless, at high energies, anomaly interactions remain dominated by pole structures that can be effectively described by light-cone-localized actions involving mixed pseudoscalar fields.\\
The implications of these results extend beyond conventional axion-like particles (ALPs) in particle physics and cosmology. They also characterize emergent ALP behavior in experimental systems such as artificial gauge fields and topological materials. In these analog systems—driven by thermal gradients and chiral chemical potentials—chiral and gravitational interactions mimic those found in anomaly physics.\\
We emphasize that local ALP interactions, due to their general couplings, can only be consistently justified in the presence of nonperturbative effects. In such cases, an anomaly pole in the ultraviolet transitions into an asymptotic state below a certain scale.

\vspace{0.3cm}
\centerline{\bf Acknowledgements}
We dedicate this work to the memory of our Colleague Elena Accomando.\\
We thank Giovanni Chirilli and Stefania D'Agostino for discussions. 
This work is partially supported by INFN, inziativa specifica {\em QG-sky}, by the the grant PRIN 2022BP52A MUR "The Holographic Universe for all Lambdas" Lecce-Naples, and by the European Union, Next Generation EU, PNRR project "National Centre for HPC, Big Data and Quantum Computing", project code CN00000013. 
\\
\\
\\
\centerline{\bf Appendix}
\appendix

\section{Absence of a particle pole for $q^2\to 0$ $(s_1,s_2,m^2)\neq 0$}
Using the Schouten relations
 \begin{align}
&p_2^{\mu_1} {\epsilon }^{\mu_2 \mu_3 p_1 p_2}= p_2^{\mu_2} {\epsilon }^{\mu_1 \mu_3 p_1 p_2}-p_2^{\mu_3} {\epsilon }^{\mu_1 \mu_2 p_1 p_2}-p_2^2 {\epsilon }^{\mu_1 \mu_2 \mu_3 p_1}+\left(  p_1 \cdot p_2\right) {\epsilon }^{\mu_1 \mu_2 \mu_3 p_2}\nonumber\\&p_1^{\mu_2} {\epsilon }^{\mu_1 \mu_3 p_1 p_2}= p_1^{\mu_1} {\epsilon }^{\mu_2 \mu_3 p_1 p_2}+p_1^{\mu_3} {\epsilon }^{\mu_1 \mu_2 p_1 p_2}-p_1^2 {\epsilon }^{\mu_1 \mu_2 \mu_3 p_2}+\left( p_1 \cdot  p_2\right) {\epsilon }^{\mu_1 \mu_2 \mu_3 p_1}
\end{align}
and by a direct perturbative computation of the vertex, we can express the $AVV$ as follows
\begin{align}
\label{nopole}
	\braket{J^{\m_1}(p_1)J^{\m_2}(p_2)J_A^{\m_3}(q)}=F_1\left( {\epsilon }^{\mu_1 \mu_2 p_1 p_2} p_1^{\mu_3} + {\epsilon }^{\mu_1 \mu_2 p_1 p_2} p_2^{\mu_3}\right)\\ \nn
	 + F_2\left({\epsilon }^{\mu_2 \mu_3 p_1 p_2} p_1^{\mu_1}-{\epsilon }^{\mu_1 \mu_2 \mu_3 p_2} s_1\right) +  F_3\left({\epsilon }^{\mu_1 \mu_3 p_1 p_2} p_2^{\mu_2}-{\epsilon }^{\mu_1 \mu_2 \mu_3 p_1} s_2\right)
\end{align}
with

\begin{align}
\label{F1}
	F_1=&\frac{1}{64 \pi ^2 \l(s,s_1,s_2)^2}\Bigl(2 \left(s^2-2 (s_1+s_2) s+(s_1-s_2)^2\right) (s-s_1-s_2)\nonumber\\&+2 \left((s_1+s_2) s^2-2 \left(s_1^2-4 s_2 s_1+s_2^2\right) s+(s_1-s_2)^2 (s_1+s_2)\right) B_0\left(s,m^2\right)\nonumber\\&-2 s_1 \left((s-s_1)^2-5 s_2^2+4 (s+s_1) s_2\right) B_0\left(s_1,m^2\right)\nonumber\\&-2 s_2 \left(s^2+4 s_1 s-2 s_2 s-5 s_1^2+s_2^2+4 s_1 s_2\right) B_0\left(s_2,m^2\right)\nonumber\\&+4 \bigl(m^2 (s-s_1-s_2) \left(s^2-2 (s_1+s_2) s+(s_1-s_2)^2\right)\nonumber\\&-s_1 s_2 \left(-2 s^2+(s_1+s_2) s+(s_1-s_2)^2\right)\bigl) C_0\left(s,s_1,s_2,m^2\right)\Bigl) 
\end{align}

\begin{align}
\label{F2}
	F_2=&\frac{1}{2 \pi ^2 \l(s,s_1,s_2)^2}\Bigl(\left(s^2-2 (s_1+s_2) s+(s_1-s_2)^2\right) (s-s_1+s_2)\nonumber\\&+s \left((s-s_1)^2-5 s_2^2+4 (s+s_1) s_2\right) B_0\left(s,m^2\right)\nonumber\\&-\left(s^3-(2 s_1+s_2) s^2+\left(s_1^2+8 s_2 s_1-s_2^2\right) s+(s_1-s_2)^2 s_2\right) B_0\left(s_1,m^2\right)\nonumber\\&+\left(-5 s^2+4 (s_1+s_2) s+(s_1-s_2)^2\right) s_2 B_0\left(s_2,m^2\right)\nonumber\\&+2 \bigl((s-s_1+s_2) \left((s-s_1)^2+s_2^2-2 (s+s_1) s_2\right) m^2\nonumber\\&+s s_2 \left(s^2+s_1 s-2 s_2 s-2 s_1^2+s_2^2+s_1 s_2\right)\bigl) C_0\left(s,s_1,s_2,m^2\right)\Bigl) 
\end{align}

\begin{align}
\label{F3}
	F_3=&-\frac{1 }{2 \pi ^2 \l(s,s_1,s_2)^2}\Bigl(\left(s^2-2 (s_1+s_2) s+(s_1-s_2)^2\right) (s+s_1-s_2)\nonumber\\&+s \left(s^2+4 s_1 s-2 s_2 s-5 s_1^2+s_2^2+4 s_1 s_2\right) B_0\left(s,m^2\right)\nonumber\\&+s_1 \left(-5 s^2+4 (s_1+s_2) s+(s_1-s_2)^2\right) B_0\left(s_1,m^2\right)\nonumber\\&-\left((s+s_1) (s-s_1)^2+(s+s_1) s_2^2-2 \left(s^2-4 s_1 s+s_1^2\right) s_2\right) B_0\left(s_2,m^2\right)\nonumber\\&+2 \bigl((s+s_1-s_2) \left((s-s_1)^2+s_2^2-2 (s+s_1) s_2\right) m^2\nonumber\\&+s s_1 \left((s-s_1)^2-2 s_2^2+(s+s_1) s_2\right)\bigl) C_0\left(s,s_1,s_2,m^2\right)\Bigl). 
\end{align}
Before performing some special kinematical limits on this amplitude, 
we comment on its UV finiteness, which is not apparent from the results for the three form factors $F_1, F_2$ and $F_3$ shown above. Notice that in these expressions the bare self-energies $B_0$ can be interpreted, indifferently, 
as either renormalized or bare. The renormalization removes the $1/\epsilon_{UV}$ poles of these integrals by defining

\beq
B_0(q^2,m^2)\to B^R_0(q^2,m^2)\equiv B_0(q^2,m^2) -\frac{1}{\epsilon_{UV}}
\eeq
One can explicitly check that by combining all such self-energy contributions and collecting all the polynomial coefficients in front, 
such ultraviolet divergences cancel, as expected. For this reason, in these expressions one can automatically replace in 
\eqref{F1}, \eqref{F2}, \eqref{F3} $B_0$ with $B^R_0$.
At this point, in order to compute the residue at $q^2=0$, given the absence of poles in the representation 
\eqref{nopole}, we need just to observe that in the expressions above for $F_1$, $F_2$ and $F_3$ no pole in $1/q^2$ are present as far as $s_1$ and $s_2$ and $m$ are nonzero. Therefore one can perform the light-cone limit, obtaining
	\begin{equation}
		\lim_{q^2\to 0} q^2 \braket{J^{\m_1}(p_1)J^{\m_2}(p_2)J^{\m_3}(q)}=0.
	\end{equation}
This result shows that there is no massless particle pole in the vertex for general kinematics. 
This result remains valid in the infrared limit of \( q \) (i.e. $q^\lambda\to 0$) as well, although the analysis simplifies significantly in this case. As already pointed out, the vanishing of the vertex in the infrared is related to the derivative coupling of 
the pseudoscalar mode, as for an ordinary NG mode.  \\

 \subsection{The continuum $\textrm{disc}\, C_0$ }
The contribution from the continuum is obtained from the s- channel cut by the ordinary cutting rules
\beq
\label{disc}
\textrm{disc}\, C_0 = \frac{1}{i \pi^2} \int d^4 k \, \frac{(-2 \pi i)^2 \delta_+(k^2 - m^2) \delta_+((k - q)^2 - m^2)}
{(k - p_1)^2 - m^2}
\eeq
that identify the light-cone character of the intermediate state.  
This integral is evaluated in the rest frame of the invariant \(q = (q_0, 0, 0, 0)\). It corresponds to a cut contribution in which the axial-vector line decays into two back-to-back on-shell fermions. This pair can be effectively treated as a pseudoscalar intermediate state, which subsequently decays into two off-shell spin-1 particles.  To evaluate the integral, we can use polar coordinates of $\vec{k}$ around the momentum $|p_1|$ together with the 
 relations 
 \begin{figure}
\label{figg2}
	\centering
	\begin{tikzpicture}
	
	\draw[thick, ->] (-1,0) -- (6,0) node[right] {$\text{Re}(s)$};
	\draw[thick, ->] (0,-3) -- (0,3) node[above] {$\text{Im}(s)$};
	
	\filldraw[black] (0,0) circle (3pt) node[below left] {$0$};
	
	\draw[thick, decorate, decoration={zigzag, amplitude=2pt, segment length=4pt}] (2,0) -- (5.9,0);
	\node at (2, -0.3) {$4m^2$};
	
		\draw[very thick, ->, black] (0,0) -- (0,-2.5); 
		\draw[very thick, ->, red] (0,0) -- (0,1.5);     

	\end{tikzpicture}
\caption{Analyticity region of the off-shell anomaly form factor in the complex \( s\equiv q^2 \)-plane, with two bold spikes at \(s=0\).  In this case there are cancelations between the cut and the contribution of the black spike, whose strength depends on all the external invariants, after integration over the $q^2>0$ region. The sum rule then allows to identify the red spike at $s=0$ as an anomaly pole.} 
\end{figure}

 \beq
 \delta_+(k^2 - m^2)=\frac{\delta( k_0- \sqrt{ |k|^2 + m^2}}{2 \sqrt{ |k|^2 + m^2} }),\qquad 
 \delta_+((k - q)^2 - 4 m^2))=\frac{1}{4 q_0  \sqrt{ q^2 -4 m^2}}\delta(|k|-\frac{1}{2} \sqrt{ q^2 -4 m^2})
 \eeq
 with 
 \beq
 k_0=\frac{\sqrt{q^2}}{2} \qquad |k|=\frac{1}{2}\sqrt{ q^2 -4 m^2}.
 \eeq
 The integral gets reduced to a simple angular form 
 \beq
 \int_{-1}^{1} \frac{d\cos\theta}{A + B \cos\theta} =\frac{1}{B} \log\left( \frac{A + B}{A-B}\right)\qquad \textrm{where} 
 \qquad  A=s_1 + s_2 - q^2, \qquad  B=2 \lambda^{1/2}(q^2,s_1,s_2)\sqrt{q^2 - 4 m^2}.
 \eeq
 Collecting all the terms together, the spectral density takes the simple form 
 \beqa
 \Delta C_0(s,s_1,s_2,m^2)\equiv \Im\, C_0(s,s_2,s_2,m^2)&=&-\frac{\pi}{4 s}\log \frac{\chi_+(s,s_1,s_2,m^2)}{\chi_-(s)}\,\theta(q^2- 4 m^2)\nonumber \\
 \chi_\pm(s,s_1,s_2,m^2) &=& s_1 + s_2 - q^2 \pm 2 \lambda^{1/2}(q^2,s_1,s_2)\sqrt{q^2 - 4 m^2}.
 \eeqa
 It is also convenient to introduce the variables	
	
	\beq
		z=\frac{1}{2 s } \left(\sqrt{\lambda} + s_1 - s_2 +s\right) \qquad \bar z=z=\frac{1}{2 s } \left(- \sqrt{\lambda} + s_1 - s_2 +s\right)
\eeq
with
	\beq
		w=\frac{1}{2} \left(1+\sqrt{1-\frac{4 m^2}{s}}\right)	\qquad \bar w=\frac{1}{2} \left(1-\sqrt{1-\frac{4 m^2}{s}}\right)
\eeq	
in order to re-express this density in the form (recall that $\Delta C_0= \textrm{disc}\, C_0/( 2 i)$)

\begin{equation}
\label{d2}
	\Delta{C_0}(s,s_1,s_2,m^2)=-\frac{  \pi  \log \left(\frac{m^2-s (w-\bar z) (\bar w-z)}{m^2-s (w-z) (\bar w-\bar z)}\right)}{s (z-\bar z)}\,\,\, \theta(s-4m^2).
\end{equation}

For the derivation of the explicit form of the spectral density  we also need $C_0$ at $s=0$  
 \beqa
 \label{d1}
C_0(0,p_1^2,p_2^2,m^2) &=& \lim_{q^2\to 0} C_0(q^2,p_1^2,p_2^2, m^2)\nonumber \\
&=&\frac{\log ^2\left(\frac{-p_1^2 +\sqrt{p_1^2  \left(p_1^2 -4 m^2\right)}+2 m^2}{2 m^2}\right)-\log ^2\left(\frac{-p_2^2 +\sqrt{p_2^2  \left(p_2^2 -4 m^2\right)}+2 m^2}{2 m^2}\right)}{2 (p_1^2 -p_2^2 )}
\eeqa
that can be inserted in the expression of the spectral density  
\begin{align}\label{ds1}
 \textrm{disc}\,\Phi_0(q^2,p_1^2,p_2^2, m^2)= -\frac{1}{2\pi^2}\biggl(&4 i \pi m^2 \, C_0 (0,p_1^2,p_2^2, m^2)\, \delta(q^2)  +2 \pi i\, \delta(q^2)\nn\\&
 -\frac{2 m^2}{q^2}\textrm{disc} \,C_0(q^2,p_1^2,p_2^2, m^2) \biggl),
\end{align}
 with the first and third contributions given respectively by Eqs. \eqref{d1} and \eqref{d2} (where $\textrm{disc} \, C_0=2 i \,\Delta C_0$). A direct evaluation of its integral in the $q^2>0$ region shows the presence of an interesting pattern of cancelations. An explicit computation, using the inverse Laplace transform, shows that the following identity holds
\beq
\label{id}
 \frac{2 i m^2}{\pi}  \, C_0 (0,p_1^2,p_2^2, m^2)=\int_{4 m^2}^\infty ds \, \rho(s,s_1,s_2,m^2)
 \eeq
where
\beqa
\label{ds2}
\rho(s,s_1,s_2,m^2)&\equiv &\frac{ m^2}{\pi^2q^2}\textrm{disc} \,C_0(q^2,p_1^2,p_2^2, m^2)\nn\\
&=&-\frac{ 2i m^2  \log \left(\frac{m^2-s (w-\bar z) (\bar w-z)}{m^2-s (w-z) (\bar w-\bar z)}\right)}{\pi s^2 (z-\bar z)}\theta(q^2 - 4 m^2),
\eeqa
 indicating that the sum rule can be thought of as being saturated by the second constant term on the rhs of \eqref{ds1}.
 Therefore, if we integrate \eqref{ds1} to compute the sum rule and use \eqref{ds2}, the first and the third term combined contribute zero to the integral, leaving only the constant pole to saturate the sum rule
 \beq
\int_0^\infty ds \,\textrm{disc}\,\Phi_0(q^2,p_1^2,p_2^2, m^2)=\frac{i}{\pi} \, \int_0^\infty ds \delta(s) = \frac{i}{\pi}.
\eeq
Further insight into this behaviour is obtained by analysing \eqref{id} under a global rescaling of all the external invariants and the mass parameter $m^2$. Notice that also in the off-shell correlator the spectral density is self-similar
\beq
\rho(\lambda s, \lambda s_1,\lambda s_2,\lambda m^2)=\frac{1}{\lambda} \rho( s, s_1, s_2,m^2)
\eeq
and that \eqref{id} is scale invariant if we perform a common rescaling of all the invariants and the mass on both of its sides.   Notice that in the on-shell limit $(s_1,s_2)\to (0,0)$ of \eqref{ref} , using in \eqref{d1}
\beq
\lim_{ (s_1,s_2)\to (0,0)} C_0(0,s_1,s_2,m^2)=-\frac{1}{2 m^2}
\eeq
we obtain 
\beq
\lim_{(s_1,s_2)\to (0,0)} \frac{2 i m^2}{\pi} \, C_0 (0,p_1^2,p_2^2, m^2)=- \frac{ i} {\pi} = \int_{4 m^2}^\infty ds \, \rho(s,m^2) 
\eeq
and we recover the sum rule that we have discussed above in the on-shell case. From this relation, if we perform the 
$m^2\to 0$ limit, we derive the spectral flow shown in Fig. 1, with  $\rho(s,m^2) $ converging to a delta function at $s=0$
\beq
\lim_{m\to 0} \rho(s,m^2)= -\frac{i}{\pi}\delta(s).
\eeq
We have seen that in the on-shell case the cancelation between the contributions of the two spikes at $s=0$, leaves the sum rule to be saturated by the cut, controlled by the mass parameter $m$. We mention again that the same result, however, can be interpreted in a different way if we perform an integration of both sides of \eqref{ds1}. After integration, the first and the third contribution 
identically cancel, as shown by \eqref{id}, and the entire sum rule appears to be saturated just by the $1/q^2$ isolated 
pole present in \eqref{zero}. In the off-shell case, a similar cancelation takes place after integration, and again, the sum rule appears to be saturated just by the pole, with a constant residue fixed only by the anomaly coefficient. \\
It is clear that we are confronted, for this interaction, with a rather special behaviour, which can be attributed to the topological properties of the vertex, not shared by other interactions.

\section{Appendix: Evaluation at the Anomalous Thresholds}
\label{app:AnomalousThresholds}

Here we collect the explicit expressions used in the cancellation analysis at the anomalous thresholds \(s = s_\pm\). These involve the evaluation of the scalar triangle integral \(C_0\) at those kinematic points.

\begin{align}
C_0(s_-,s_1,s_2,m^2) &= \frac{1}{{s_1}{s_2}\, s_-^2}\Biggl(\sqrt{{s_1}} \Biggl(\sqrt{{s_2} \left({s_2}-4 m^2\right)} \left(\sqrt{{s_1}}-\sqrt{{s_2}}\right)\nn\\
&\times \log \left(\frac{\sqrt{{s_2} \left({s_2}-4 m^2\right)}+2 m^2-{s_2}}{2 m^2}\right)\nonumber\\
&+\sqrt{{s_2}} \sqrt{\left(-2 \sqrt{{s_1}} \sqrt{{s_2}}+{s_1}+{s_2}\right) \left(-4 m^2-2 \sqrt{{s_1}} \sqrt{{s_2}}+{s_1}+{s_2}\right)} \log \left(S_1\right)\Biggl)\nonumber\\
&+\sqrt{{s_2}} \sqrt{{s_1} \left({s_1}-4 m^2\right)} \left(\sqrt{{s_2}}-\sqrt{{s_1}}\right) \log \left(\frac{\sqrt{{s_1} \left({s_1}-4 m^2\right)}+2 m^2-{s_1}}{2 m^2}\right)\Biggl)
\end{align}

\begin{align}
	C_0(s_+,s_1,s_2,m^2) &= \frac{1}{{s_1}{s_2} s_+^2}\Biggl(\sqrt{{s_1}} \Biggl(\sqrt{{s_2} \left({s_2}-4 m^2\right)} \left(\sqrt{{s_1}}+\sqrt{{s_2}}\right)\nn\\
	&\times \log \left(\frac{\sqrt{{s_2} \left({s_2}-4 m^2\right)}+2 m^2-{s_2}}{2 m^2}\right)\nonumber\\
	&-\sqrt{{s_2}} \sqrt{\left(\sqrt{{s_1}}+\sqrt{{s_2}}\right)^4-4 m^2 \left(\sqrt{{s_1}}+\sqrt{{s_2}}\right)^2} \log \left(S_2\right)\Biggl)\nonumber\\
	&+\sqrt{{s_2}} \sqrt{{s_1} \left({s_1}-4 m^2\right)} \left(\sqrt{{s_1}}+\sqrt{{s_2}}\right) \log \left(\frac{\sqrt{{s_1} \left({s_1}-4 m^2\right)}+2 m^2-{s_1}}{2 m^2}\right)\Biggl),
\end{align}

where
\begin{align}
	S_1 &= \frac{\sqrt{\left(-2 \sqrt{{s_1}} \sqrt{{s_2}}+{s_1}+{s_2}\right) \left(-4 m^2-2 \sqrt{{s_1}} \sqrt{{s_2}}+{s_1}+{s_2}\right)}+2 m^2-\left(\sqrt{{s_1}}-\sqrt{{s_2}}\right)^2}{2 m^2}, \\
	S_2 &= \frac{\sqrt{\left(\sqrt{{s_1}}+\sqrt{{s_2}}\right)^4-4 m^2 \left(\sqrt{{s_1}}+\sqrt{{s_2}}\right)^2}+2 m^2-\left(\sqrt{{s_1}}+\sqrt{{s_2}}\right)^2}{2 m^2}.
\end{align}

Using these expressions, one can show:
\[
\Phi_{\text{ATT}}^{(\lambda)}(q^2 = s_\pm) = 0.
\]

\subsection{The massless and the on-shell limits do not commute}
An interesting limit of the amplitude arises when either the fermion mass or the external invariants \( s_1 \) and \( s_2 \) are sent to zero. As evident from our previous discussions, these two limits do not commute for an obvious reason. If the on-shell limit on \( s_1 \) and \( s_2 \) is taken first, the interaction is reduced to its longitudinal sector. As shown above, the sum rule then causes the branch cut to collapse into a pole when approaching the conformal limit by setting \( m \to 0 \). In this case, the anomaly form factor becomes the only surviving form factor in the vertex, with a nonzero residue as \( q^2 \to 0 \), equaling the numerical coefficient of the anomaly.
Conversely, if we first take the \( m \to 0 \) limit while keeping the vector lines off-shell, and only afterward send \( q^2 \to 0 \), the residue vanishes. This happens because, in this order of limits, both the longitudinal and transverse structures contribute. In the following, we will illustrate this point in detail.
\\
We start from the $m=0$ case. To show the patterns of cancelations we assume that the two external vector lines are off-shell. 
\begin{itemize}
	\item  $m=0$\\
	In the $m=0$ case, a direct computation of the residue, combining all the terms in \eqref{nopole} gives 
	\begin{equation}
	\label{res}
		q^2 \braket{J^{\m_1}(p_1)J^{\m_2}(p_2)J^{\m_3}(q)}=-\frac{   \left(\text{B}^R_0(M^2,0)+M^2 C_0(0,M^2,M^2,0)-2\right) q^{\m_3}\epsilon^{\m_1 \m_2 p_1 p_2}}{8 \pi ^2}
	\end{equation}
	where we compute the scalar 3-point function at $q^2=0$ (with  $s_1=s_2=M^2 <0)$
	\begin{equation}
		C_0(0,M^2,M^2,0) =\lim_{q^2\to 0} C_0(q^2,M^2,M^2,0)= -\frac{1}{\varepsilon_{IR}  M^2}-\frac{\log \left(-\frac{\pi  \mu ^2}{M^2}\right)}{M^2}+\frac{\gamma }{M^2}+\frac{2 \log (\pi )}{M^2}
	\end{equation}
	as well as the scalar 2-point function $B_0^R$
		\begin{equation}
		\text{B}^R_0(0,M^2)\equiv \lim_{q^2\to 0} B_0^R(q^2 M^2)= \log \left(-\frac{\mu ^2}{\pi  M^2}\right)-\gamma +2,
	\end{equation}
	with $\varepsilon= d-4$ and $\gamma$ being the Euler-Mascheroni constant and $\mu$ an infrared renormalization scale.  
	\eqref{res} contains a light-cone component $q^{\mu_3}\sim q^+ n^{+\, \mu_3}$, that remains large but fixed in the 
	light-cone limit that we are considering.  At this stage, the vanishing of the residue 
	\beq
	\lim_{q^2\to 0} \left(\text{B}^R_0(q^2, M^2)+M^2 C_0(q^2,M^2,M^2)-2\right) =0
	\eeq
	 eliminates such component. 
	 Notice that there are no infrared divergences in $1/\varepsilon_{IR}$ in this limit.
	Therefore we derive the relation
	 \begin{equation}
	\lim_{q^2\to 0}	q^2 \braket{J^{\m_1}(p_1)J^{\m_2}(p_2)J^{\m_3}(q)}=0,
	\end{equation}
	proving the absence of any particle pole in the vertex under these external kinematical constraints.
	
	\item  $m\neq 0$\\
	A similar analysis can be performed for a nonzero mass. The result is finite in the $q^2\to 0$ limit. We obtain
	\begin{align}
		q^2 \braket{J^{\m_1}(p_1)J^{\m_2}(p_2)J^{\m_3}(q)}=\frac{1  }{8 \pi ^2}q^{\m_3}\epsilon^{\m_1 \m_2 p_1 p_2} \bigl(&-\text{B}^R_0\left(M^2,m^2\right)+\text{B}^R_0\left(0,m^2\right)\nn\\&+\left(4 m^2-M^2\right) C_0\left(0,M^2,M^2,m^2\right)+2\bigl)
	\end{align}
	which is purely longitudinal due to the presence of the $q^{\m_3}\epsilon^{\m_1 \m_2 p_1 p_2}$ tensor,
	where
	\begin{equation}
		\text{B}^R_0\left(M^2,m^2\right)= \log \left(\frac{\mu ^2}{\pi  m^2}\right)+\frac{\sqrt{M^2 \left(M^2-4 m^2\right)} \log \left(\frac{\sqrt{M^2 \left(M^2-4 m^2\right)}+2 m^2-M^2}{2 m^2}\right)}{M^2}-\gamma +2,
	\end{equation}
		\begin{equation}
		\text{B}^R_0\left(0,m^2\right)= \log \left(\frac{\mu ^2}{\pi  m^2}\right)-\gamma,
	\end{equation}
	and
	\begin{equation}
		C_0\left(0,M^2,M^2,m^2\right)=\frac{\sqrt{-\left(M^2 \left(4 m^2-M^2\right)\right)} \log \left(\frac{\sqrt{M^2 \left(M^2-4 m^2\right)}+2 m^2-M^2}{2 m^2}\right)}{M^2 \left(4 m^2-M^2\right)}.
	\end{equation}
Also in this case there is no particle pole since
\beq
\lim_{q^2\to 0} \left( -\text{B}^R_0\left(M^2,m^2\right)+\text{B}^R_0\left(q^2,m^2\right)+\left(4 m^2-M^2\right) 
C_0\left(q^2,M^2,M^2,m^2\right)+2 \right)=0
\eeq
giving 
		\begin{align}
	\lim_{q^2\to 0} 	q^2 \braket{J^{\m_1}(p_1)J^{\m_2}(p_2)J^{\m_3}(q)}=0.
	\end{align}
\end{itemize}
The analysis shows that the transverse sector turns into a longitudinal one when we take the light-cone limit, a fact previously noted in \cite{Armillis:2009sm}. \\
Starting from the general expression \eqref{res}, setting the vector lines on-shell and keeping the mass parameter fixed, results in the absence of a particle pole in the vertex. This is consistent with the observation that, in this case, there is no pole but rather a cut in the longitudinal form factor. 
Therefore, even in the on-shell case (\(s_1 = s_2 = 0\)), where the vertex is purely longitudinal, for \(m \neq 0\), we obtain
	
\begin{align}
\label{zzz}
		q^2 \braket{J^{\m_1}(p_1)J^{\m_2}(p_2)J^{\m_3}(q)}=&\frac{ \left(2 m^2 C_0\left(0,0,0,m^2\right)+1\right) q^{\m_3} \epsilon^{\m_1 \m_2 p_1 p_2}}{2 \pi ^2}
	\end{align}
which has no pole since, according to \eqref{exp1} 
	\begin{equation}
		C_0\left(0,0,0,m^2\right)=-\frac{1}{2m^2}
	\end{equation}
and \eqref{zzz} vanishes. The sum rule, however, indicates that as $m\to 0$, the spectral density and the area law 
of the sum rule 	generate a pole, as in the conformal analysis. At the conformal point the form factor with its longitudinal tensor structure equals the entire vertex   
	\begin{equation}
		\lim_{q^2\to 0} q^2 \braket{J^{\m_1}(p_1)J^{\m_2}(p_2)J^{\m_3}(q)}=\frac{1 } {2 \pi ^2} q^{\m_3} \epsilon^{\m_1 \m_2 p_1 p_2}
	\end{equation}
showing a nonzero residue.\
This demonstrates that the vertex possesses a particle pole with a nonvanishing coupling \( g_{pp} \), as defined in \eqref{loc} for the general case, at \( s = 0 \). This occurs only for on-shell vector lines and a massless fermion. Notice that the limit we are considering is not a soft limit, and clearly characterizes a process that is entirely supported on the light-cone.\\
On the other end, in the infrared case, as \( q^\lambda \to 0 \), multiplication by the invariant \( q^2 \) leads to a contribution of the form \( \epsilon^{\mu \nu p_1 p_2} \). These contributions vanish due to symmetry, given that \( p_1 + p_2 \sim 0 \). Consequently, one deduces that no particle pole exists in a general sense, except under specific kinematical conditions. While the anomaly form factor is characterized by an anomaly pole, as shown through the analysis of the conformal Ward identities, this does not translate into a general particle pole. \\
This indicates that topological vertices cannot be directly assimilated into ordinary, non-topological ones. The observed cancellations between the longitudinal and transverse structures of the interaction lead to this conclusion. Moreover, the non-commutativity between the massless and on-shell limits is a property of the entire vertex, stemming from the interrelation between these two sectors. 
These analysis show that the anomaly pole behaves as a particle pole only in the on-shell and massless case where
\begin{equation}
		 \braket{J^{\m_1}(p_1)J^{\m_2}(p_2)J^{\m_3}(q)}=\frac{1} {2 \pi ^2 } \frac{q^{\m_3}}{q^2} \epsilon^{\m_1 \m_2 p_1 p_2}
	\end{equation} 
is present over the entire light-cone, getting arbitrarily soft around the apex of the cone where $q^\lambda\to 0$.  

\section{The transverse form factors of the $ATT$}
\label{transverse}
\small
\begin{equation}
	\begin{aligned}
		\bar{A}_{11}=&-2 s_1 \left\{2 \lambda m^2  \left[ 21 s^3 - 3 s^2 (8 s_1 + 7 s_2) + s \left( -15 s_1^2 + 100 s_1 s_2 - 21 s_2^2 \right) + 3 (s_1 - s_2)^2 (6 s_1 + 7 s_2) \right] \right.\\&
		+ \left. s_1 \left[ 3 s^5 + s^4 (81 s_2 - 12 s_1) + 2 s^3 (9 s_1^2 - 40 s_1 s_2 - 42 s_2^2) - 4 s^2 (3 s_1^3 + 20 s_1^2 s_2 - 76 s_1 s_2^2 + 21 s_2^3) + 3 s_2 (s_1 - s_2)^4\right.\right.\\&
		+ \left.\left. s (s_1 - s_2)^2 (3 s_1^2 + 82 s_1 s_2 + 81 s_2^2)  \right]\right\}, \\
		\bar{A}_{12}=&-4 \lambda m^2 \Big[-3 s_2^2 (s^2+14 s s_1-17 s_1^2)-3 s_2 (s-s_1) (s^2-21 s s_1-6 s_1^2)+s_2^3 (7 s-24 s_1)+2 (s-s_1)^3 (s+3 s_1)-3 s_2^4\Big]\\&+2 s_2 \Big[s^6-s^5 (14 s_1+5 s_2)+s^4 (-11 s_1^2+28 s_1 s_2+10 s_2^2)+2 s^3 (52 s_1^3-102 s_1^2 s_2-5 s_2^3)+s^2 (-121 s_1^4+152 s_1^3 s_2+\\&120 s_1^2 s_2^2-28 s_1 s_2^3+5 s_2^4)+s (s_1-s_2)^2 (38 s_1^3+117 s_1^2 s_2+12 s_1 s_2^2-s_2^3)+3 s_1^2 (s_1-s_2)^4\Big],\\
		\bar{A}_{13}=&4 \lambda m^2 \Big[3 s^4+s^3 (24 s_1-7 s_2)+3 s^2 (-17 s_1^2+14 s_1 s_2+s_2^2)+3 s (s_1-s_2) (6 s_1^2+21 s_1 s_2-s_2^2)+2 (s_1-s_2)^3 (3 s_1+s_2)\Big]\\&+2 s \Big[-2 s_2^3 (5 s^3+102 s s_1^2-52 s_1^3)+s_2^4 (10 s^2+28 s s_1-11 s_1^2)-s_2 (s-s_1)^2 (s^3-12 s^2 s_1-117 s s_1^2-38 s_1^3)\\&+s_2^2 (5 s^4-28 s^3 s_1+120 s^2 s_1^2+152 s s_1^3-121 s_1^4)+3 s_1^2 (s-s_1)^4-s_2^5 (5 s+14 s_1)+s_2^6\Big],\\
		\bar{A}_{14}=&24 \Big\{ \lambda^2 m^4 \Big[s^2+s (s_1-2 s_2)-2 s_1^2+s_1 s_2+s_2^2\Big]+m^2 s_1 \Big[2 s_2^3 (16 s^3-15 s s_1^2-5 s_1^3)-18 s s_2^2 (s^3+s s_1^2-2 s_1^3)-9 s_1 s_2^5\\&+2 s_2^6+3 s_2^4 (-6 s^2+3 s s_1+5 s_1^2)+3 s_1 s_2 (s_1-3 s) (s_1-s)^3+(s-s_1)^5 (2 s+s_1)\Big]+s s_1^2 s_2 \Big[3 s_2^2 (-4 s^2+4 s s_1+s_1^2)\\&+3 s_2 (s^3+4 s^2 s_1-6 s s_1^2+s_1^3)+s_2^3 (3 s-7 s_1)+(s-s_1)^3 (3 s+2 s_1)+3 s_2^4\Big]\Big\},\\
		\bar{A}_{15}=&\lambda \Big[4 \lambda m^2 (5 s^2+13 s s_1-10 s s_2-18 s_1^2+13 s_1 s_2+5 s_2^2)-2 s_2^3 (s^2-8 s s_1+8 s_1^2)-(s-s_1)^3 (s^2-5 s s_1-2 s_1^2)\\&-2 s_2^2 (s^3+24 s^2 s_1-38 s s_1^2-5 s_1^3)+s_2 (s-s_1) (3 s^3+19 s^2 s_1+95 s s_1^2-s_1^3)+s_2^4 (3 s+8 s_1)-s_2^5\Big]
	\end{aligned}
\end{equation}

\normalsize
and
\small
\begin{equation}
	\begin{aligned}
		\bar{A}_{21}=&  4 s_1 
		\Big[ s \, s_1 \Big( s_2^2 ( 9 s - 19 s_1 ) - 2 s_2 ( 9 s + 4 s_1 ) (s - s_1) - (s - s_1)^3 + 10 s_2^3 \Big) - \lambda m^2 \Big(
		17 s^2 - 8 s (s_1 + s_2) - 9 (s_1 - s_2)^2 \Big) \Big],\\ 
		\bar{A}_{22}=& 2 s s_2 \Big[s^4 - 2 s^3 (5 s_1 + 2 s_2) + 6 s^2 s_2 (2 s_1 + s_2) + 
		s (26 s_1^3 - 60 s_1^2 s_2 + 6 s_1 s_2^2 - 4 s_2^3) - (s_1 - 
		s_2)^2 (17 s_1^2 + 6 s_1 s_2 - s_2^2)\Big]\\& - 
		4 \lambda m^2 \Big[-s_2 (s^2 - 30 s s_1 + 9 s_1^2) + 
		2 (s - s_1)^2 (s + 3 s_1) - 4 s s_2^2 + 
		3 s_2^3\Big] ,\\
		\bar{A}_{23}=& 2 
		s \Big[ 
		2 s^3 ( s_1^2 + 5 s_1 s_2 + 2 s_2^2 ) - 6 s^2 ( 
		s_1^3 - 6 s_1^2 s_2 + 2 s_1 s_2^2 + s_2^3 ) + 2 s (s_1 - s_2) \big( 
		3s_1^3-20s_1^2s_2+s_1s_2^2-2s_2^3
		\big) \\&-(s_1-s_2)^3(2s_1^2+5s_1s_2-s_2^2) -s_2 s^4
		\Big] +
		4 \lambda m^2 \Big[
		3 s^3 + 2 s^2 (9 s_1 - 2 s_2)  - 2 (s_1 - s_2)^3 - s (s_1 - s_2) ( 
		19 s_1 - s_2 )
		\Big] ,\\\bar{A}_{24}=& 
		12 \big( \lambda m^2 + s \, s_1 \, s_2 \big) \Big[ 
		2 \lambda m^2 (s + s_1 - s_2) + s_1 \Big( 
		3 s^3 + s^2 (3 s_2 - 5 s_1) + s (s_1 - s_2) (s_1 + 5 s_2) + (s_1 - s_2)^3 
		\Big) \Big] , \\
		\bar{A}_{25}=& \lambda \Big[20 \lambda m^2 (s + s_1 - s_2) - s^4 + 
		2 s^3 (3 s_1 + s_2) + 4 s^2 s_1 (3 s_2 - 2 s_1) + 
		2 s (s_1 - s_2) (s_1^2 + 8 s_1 s_2 + s_2^2) + (s_1 - s_2)^4\Big].
	\end{aligned}
\end{equation}
\normalsize

\providecommand{\href}[2]{#2}\begingroup\raggedright\endgroup

\end{document}